\documentclass[11pt]{amsart}%
\usepackage{graphicx}
\usepackage{amscd}
\usepackage{amsbsy}
\usepackage{amsmath}
\usepackage{amstext}%
\usepackage{amsfonts}%
\usepackage{amssymb}
\newtheorem{theorem}{Theorem}
\theoremstyle{plain}

\newtheorem{definition}{Definition}
\newtheorem{example}{Example}

\newtheorem{proposition}{Proposition}
\newtheorem{remark}{Remark}

\numberwithin{equation}{section}

\makeindex

\begin{document}
\title[A Tangled Tale of Quantum Entanglement]{An Entangled Tale of Quantum Entanglement\\Version 1.5}
\author{Samuel J. Lomonaco, Jr.}
\address{Dept. of Comp. Sci. \& Elect. Engr.\\
University of Maryland Baltimore County\\
1000 Hilltop Circle\\
Baltimore, MD 21250}
\email{E-Mail: Lomonaco@UMBC.EDU}
\urladdr{WebPage: http://www.csee.umbc.edu/\symbol{126}lomonaco}
\thanks{This work was partially supported by ARO Grant \#P-38804-PH-QC, NIST, and the
L-O-O-P Fund. The author gratefully acknowledges the hospitality of the
University of Cambridge Isaac Newton Institute for Mathematical Sciences,
Cambridge, England, where some of this work was completed. \ Thanks are also
due to the other AMS Short Course lecturers, Howard Brandt, Dan Gottesman, Lou
Kauffman, Alexei Kitaev, Peter Shor, Umesh Vazirani and the many Short Course
participants for their support. (Copyright 2001)}
\date{December 30, 2000}

\begin{abstract}
These lecture notes give an overview from the perspective of Lie group theory
of some of the recent advances in the rapidly expanding research area of
quantum entanglement. \ 

This paper is a written version of the last of eight one hour lectures given
in the American Mathematical Society (AMS)\ Short Course on Quantum
Computation held in conjunction with the Annual Meeting of the AMS in
Washington, DC, USA in January 2000.

More information about the AMS\ Short Course can be found at the website:
http://www.csee.umbc.edu/\symbol{126}lomonaco/ams/Announce.html

\end{abstract}
\maketitle
\tableofcontents

\section{Introduction}

\bigskip

These lecture notes were written for the American Mathematical Society (AMS)
Short Course on Quantum Computation held 17-18 January 2000 in conjunction
with the Annual Meeting of the AMS in Washington, DC in January 2000. \ \bigskip

The objective of this lecture is to discuss quantum entanglement from the
perspective of the theory of Lie groups. \ More specifically, the ultimate
objective of this paper is to quantify quantum entanglement in terms of Lie
group invariants, and to make this material accessible to a larger audience
than is currently the case. \ These notes depend extensively on the material
presented in Lecture I \cite{Lomonaco1}. \ It is assumed that the reader is
familiar with the material on density operators and quantum entanglement given
in the AMS\ Short Course Lecture I, i.e., with sections 5 and 7 of
\cite{Lomonaco1}. \ 

\bigskip

Of necessity, the scope of this paper is eventually restricted to the study of
qubit quantum systems, and to a specific problem called the \emph{Restricted
Fundamental Problem in Quantum Entanglement} (\emph{RFPQE}). \ References to
the broader scope of quantum entanglement are given toward the end of the paper.

\bigskip

\subsection{Preamble}%

\begin{center}
\fbox{\includegraphics[
height=4.1753in,
width=2.9724in
]%
{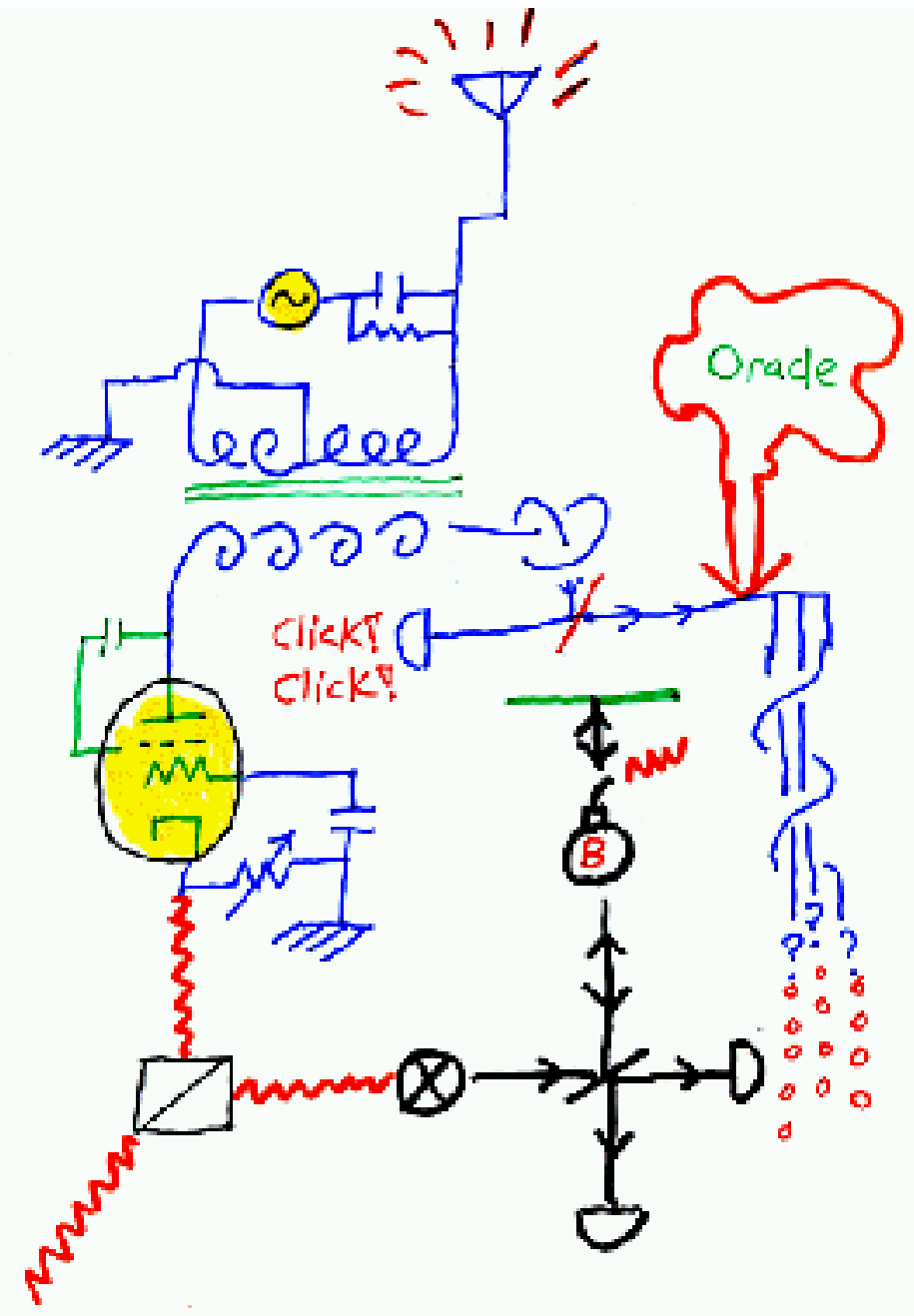}%
}\\
Figure 1. \ Quantum Entanglement Lab???
\end{center}
\bigskip

At first sight, a physics research lab dedicated to the pursuit of quantum
entanglement might look something like the drawing found in Figure 1, i.e.,
like an indecipherable, incoherent jumble of wires, fiber optic cable, lasers,
bean splitters, lenses. \ Perhaps some large magnets for NMR equipment, or
some supercooling equipment for rf SQUIDs are tossed in for good measure.
\ Whatever .. It is indeed a most impressive collection of adult ``toys.''

\bigskip

However, to a mathematician, such a lab appears very much like a well
orchestrated collection of intriguing mathematical ``toys,''\ just beckoning
with new tantalizing mathematical challenges. \ 

\bigskip

\subsection{A Sneak Preview}

\qquad\bigskip

In the hope of piquing your curiosity to read on, we give the following brief
preview of what is to come:

\bigskip

The RFPQE reduces to the mathematical problem of determining the orbits of the
big adjoint action of the group of local unitary transformations
$\mathbb{L}\left(  2^{n}\right)  $ on the Lie algebra $u\left(  2^{n}\right)
$ of the unitary group $\mathbb{U}\left(  2^{n}\right)  $, as expressed by the
following formula:
\[
\fbox{$\mathbb{L}\left(  2^{n}\right)  \times u(2^{n})\overset{Ad}%
{\longrightarrow}u(2^{n})$}%
\]
where ``$Ad$'' denotes the big adjoint operator, and where the remaining
symbols are defined in the table below.
\[%
\begin{tabular}
[c]{|c||c|}\hline
$\mathbb{L}\left(  2^{n}\right)  =%
{\displaystyle\bigotimes\limits_{1}^{n}}
\mathbb{SU}\left(  2\right)  $ & Local Unitary Group\\\hline
$\ell\left(  2^{n}\right)  =\overset{n}{\underset{1}{\boxplus}}su\left(
2\right)  $ & Lie Algebra of $\mathbb{L}\left(  2^{n}\right)  $\\\hline
$\mathbb{U}\left(  2^{n}\right)  $ & Unitary Group\\\hline
$u\left(  2^{n}\right)  $ & Lie Algebra of $\mathbb{U}\left(  2^{n}\right)
$\\\hline
\end{tabular}
\]

\bigskip

We attack this problem by lifting the above big adjoint action to the
\textbf{induced infinitesimal action}
\[
\fbox{$\ell(2^{n})\overset{\Omega}{\longrightarrow}\mathbf{Vec}((\mathbf{u}%
(2^{n}))$}%
\]
which, for a $3$ qubit density operator $\rho$, is explicitly given
by\footnote{This expression will be explained later in the paper. \ I hope
that this will make you curious enough to read on?}
\[
\fbox{$%
\begin{array}
[c]{c}%
\Omega\left(  v\right)  \left(  i\rho\right)  =%
{\displaystyle\sum\limits_{q_{1},q_{2}=0}^{3}}
\left(  a^{(1)}\cdot x_{\ast q_{1}q_{2}}\times\frac{\partial}{\partial x_{\ast
q_{1}q_{2}}}+a^{(2)}\cdot x_{q_{1}\ast q_{2}}\times\frac{\partial}{\partial
x_{q_{1}\ast q_{2}}}+a^{(3)}\cdot x_{q_{1}\ast q_{2}}\times\frac{\partial
}{\partial x_{q_{1}q_{2}\ast}}\right)
\end{array}
$}%
\]

\bigskip

\noindent where $v\in\mathbf{\ell}\left(  2^{3}\right)  $ and $i\rho
\in\mathbf{u}\left(  2^{3}\right)  $ are given by
\[
\left\{
\begin{array}
[c]{ccl}%
v & = & a^{(1)}\cdot\xi_{\ast00}+a^{(2)}\cdot\xi_{0\ast0}+a^{(3)}\cdot
\xi_{00\ast}\\
&  & \\
i\rho & = &
{\displaystyle\sum\limits_{r_{1},r_{2},r_{3}=0}^{3}}
x_{r_{1}r_{2}r_{3}}\xi_{r_{1}r_{2}r_{3}}%
\end{array}
\right.
\]
and where $\mathbf{Vec}((\mathbf{u}(2^{n}))$ denotes the Lie algebra of vector
fields on $\mathbf{u}\left(  2^{n}\right)  $.

\bigskip

The induced infinitesimal action can then be used to quantify and to classify
quantum entanglement through the construction of a complete set of quantum
entanglement invariants.

\bigskip

In the pages to follow, we make every effort to make the above sneak preview
more transparent and understandable. \ Our goal is to present the underlying
intuitions without getting lost in an obscure haze of technicalities.
\ However, presenting this topic is much like tiptoeing through a mine field.
\ One false move, and everything explodes into a dense jungle and clutter of
technicalities. \ We leave it to the reader to determine how successful this
endeavor is.

\bigskip

\subsection{How our view of quantum entanglement has dramatically changed over
this past century}

\qquad\bigskip

Finally, we close this introduction with a brief historical perspective.

\bigskip

Over the past twentieth century, the scientific community's view of quantum
entanglement has dramatically changed. \ It continues to do so even today.

\bigskip

Initially, quantum entanglement was viewed as an unnecessary and unwanted wart
on quantum mechanics. \ Einstein, Podolsky, and Rosen\cite{Einstein1} tried to
surgically remove it. \ Bell\cite{Bell1},\cite{Bell2} showed that such surgery
can not be performed without destroying the very life of physical reality.

\bigskip

Today, quantum entanglement is viewed as a useful resource within quantum
mechanics. \ It is now viewed as a commodity to be utilized and traded, much
as would be a commodity on the stock exchange.

\bigskip

Quantum entanglement appears to be one of the physical phenomena at the
central core of quantum computation. \ Many believe that it is quantum
entanglement that somehow enables us to harness the vast parallelism of
quantum superposition.

\bigskip

But what is quantum entanglement?

\bigskip

How do we measure, quantify, classify quantum entanglement? \ When is the
quantum entanglement of two quantum systems the same? different? \ When is the
quantum entanglement of one quantum system greater than that of another?

\bigskip

It is anticipated that answers to the above questions will have a profound
impact on the development of quantum computation. \ Finding answers to these
questions is challenging, intriguing, and indeed very habit forming.

\bigskip

\section{A Story of Two Qubits, or How Alice \& Bob Learn to Live with Quantum
Entanglement \ and Love It.}

\bigskip

Our tangled tale of quantum entanglement begins with Alice and Bob's first
encounter with quantum entanglement.

\bigskip

Alice and Bob, who happen to be good friends (as attested, time and time
again, by the open literature on quantum computation), meet one day. \ A
discussion ensues. \ The topic, of course, is quantum entanglement.
\ Fortunately or unfortunately, depending on how one looks at it, their
discussion explodes into a heated argument. \ After a lengthy debate, they
agree that the only way to resolve their conflict is to purchase the real
McCoy, i.e., a pair of entangled qubits. \ So they rush to the nearest
\textsc{Toys for Aging Children} \textsc{Store} to see what they can find. \ 

\bigskip

Almost immediately upon entering the store, they happen to spy, on one of the
store shelves, an elaborately decorated box labelled:

\bigskip%

\[
\fbox{%
\begin{tabular}
[c]{c}%
$\mathbb{Q}$\textbf{.}$\mathbb{E}$\textbf{., Inc.}\\\hline\hline
$\text{\textbf{Two Entangled Qubits}}$\\
$\mathcal{Q}_{AB}$\\
$\text{Consisting of qubits}$\\
$\mathcal{Q}_{A}\text{ and }\mathcal{Q}_{B}$%
\end{tabular}
}%
\]
On the back of the box is the content label, required by federal law, which reads:

\bigskip%

\[%
\begin{tabular}
[c]{|ccccc|}\hline
&  & $%
\begin{array}
[c]{c}%
\text{{\scriptsize U.S. Certified}}\\
\text{\textbf{Contents}}\\
\quad\fbox{$%
\begin{array}
[c]{c}%
\text{\textsf{EPR Pair}}%
\end{array}
$}^{(\ast)}%
\end{array}
$ &  & \\\hline\hline
\textbf{Q Sys} & \multicolumn{1}{|c}{$%
\begin{array}
[c]{c}%
\text{\textbf{Hilb.}}\\
\text{\textbf{Sp.}}%
\end{array}
$} & \multicolumn{1}{|c}{\textbf{State}} & \multicolumn{1}{|c}{$%
\begin{array}
[c]{c}%
\text{\textbf{Unitary}}\\
\text{\textbf{Transf.}}%
\end{array}
$} & \multicolumn{1}{|c|}{$%
\begin{array}
[c]{c}%
\text{\textbf{State}}\\
\text{\textbf{Space}}%
\end{array}
$}\\\hline\hline
$\mathcal{Q}_{AB}$ & \multicolumn{1}{|c}{$\mathcal{H}_{AB}$} &
\multicolumn{1}{|c}{$\rho_{AB}=\overset{}{\underset{}{\left(
\begin{array}
[c]{rrrr}%
\frac{1}{2} & 0 & 0 & -\frac{1}{2}\\
0 & 0 & 0 & 0\\
0 & 0 & 0 & 0\\
-\frac{1}{2} & 0 & 0 & \frac{1}{2}%
\end{array}
\right)  }}$} & \multicolumn{1}{|c}{$\mathbb{U}(2^{2})_{AB}$} &
\multicolumn{1}{|c|}{$u(2^{2})_{AB}$}\\\hline
$\mathcal{Q}_{A}$ & \multicolumn{1}{|c}{$\mathcal{H}_{A}$} &
\multicolumn{1}{|c}{$\rho_{A}=\overset{}{\underset{}{\left(
\begin{array}
[c]{cc}%
\frac{1}{2} & 0\\
0 & \frac{1}{2}%
\end{array}
\right)  }}$} & \multicolumn{1}{|c}{$\mathbb{U}(2)_{A}$} &
\multicolumn{1}{|c|}{$u(2)_{A}$}\\\hline
$\mathcal{Q}_{B}$ & \multicolumn{1}{|c}{$\mathcal{H}_{B}$} &
\multicolumn{1}{|c}{$\rho_{B}=\overset{}{\underset{}{\left(
\begin{array}
[c]{cc}%
\frac{1}{2} & 0\\
0 & \frac{1}{2}%
\end{array}
\right)  }}$} & \multicolumn{1}{|c}{$\mathbb{U}(2)_{B}$} &
\multicolumn{1}{|c|}{$u(2)_{B}$}\\\hline
&  &
\begin{tabular}
[c]{l}%
{\tiny (*) Caveat Emptor: \ \ Not legally responsible}\\
\quad{\tiny \qquad\qquad\qquad\qquad\ for the effects of decoherence.}%
\end{tabular}
&  & \\\hline
\end{tabular}
\]

\bigskip

Alice and Bob hurriedly purchase the two qubit quantum system $\mathcal{Q}%
_{AB}$. \ Outside the store, they rip open the box. \ Alice grabs the qubit
labelled $\mathcal{Q}_{A}$. \ Bob then takes the remaining qubit
$\mathcal{Q}_{B}$.

\bigskip

Alice and Bob then immediately\footnote{For some unknown reason, everyone
involved with the quantum world is always in a hurry. \ Perhaps such haste is
caused by concerns in regard to decoherence?} depart for their separate
destinations. \ Alice flies to Queensland, Australia to continue with her
Ph.D. studies at the University of Queensland. \ She arrives just in time to
attend the first class lecture on quantum mechanics. \ Bob, on the other hand,
flies to Vancouver, British Columbia to continue with his Ph.D. studies at the
University of British Columbia. \ He just barely arrives in time to hear the
first lecture in a course on differential geometry and Lie groups.

\bigskip

Soon after her quantum mechanics lecture, Alice begins to have second thoughts
about their joint purchase of two entangled qubits. \ She quickly reaches for
her cellphone, calls Bob, and nervously fires off in rapid succession three questions:

\begin{quotation}
``Did we get our money's worth of quantum entanglement?''

\noindent``How much quantum entanglement did we actually purchase?''

\noindent``Are we the victims of a modern day quantum entanglement scam?''
\end{quotation}

\bigskip

After the phone conversation, Bob is indeed deeply concerned. \ In
desperation, he calls the U.S. Quantum Entanglement Protection Agency, which
refers him to the U.S. National Institute of Quantum Entanglement Standards
and Technology (NI$_{\mathbb{QE}}$ST) in Gaithersburg, Maryland.

\bigskip

After a long conversation, a representative of NI$_{\mathbb{QE}}$ST agrees to
send Alice and Bob, free of charge, the NI$_{\mathbb{QE}}$ST Quantum
Entanglement Standards Kit. \ On hanging up, the NI$_{\mathbb{QE}}$ST
representative takes the NI$_{\mathbb{QE}}$ST standard entangled two qubit
quantum system $\mathcal{Q}_{A^{\prime}B^{\prime}}^{\prime}$ off the shelf,
places $\mathcal{Q}_{A^{\prime}}^{\prime}$ together with a User's Manual into
a box marked ``Alice.'' \ He/She also places the remaining qubit
$\mathcal{Q}_{B^{\prime}}^{\prime}$ together with a User's Manual into a
second box labeled ``Bob,'' \ and then sends the two boxes by overnight mail
to Alice and Bob respectively.

\bigskip

The very next day (in different time zones, of course) Alice and Bob each
receive their respective packages, take out their respective qubits, and read
the enclosed user's manuals.

\bigskip

The NI$_{\mathbb{QE}}$ST User's Manual reads as follows:

\begin{quotation}
\bigskip

\noindent$\mathbb{Q}$\textsc{.}$\mathbb{E}$\textsc{. Yardstick 1}\textbf{.}
\ An EPR pair $\mathcal{Q}_{AB}$ possess the same quantum entanglement as the
NI$_{\mathbb{QE}}$ST standard EPR pair $\mathcal{Q}_{A^{\prime}B^{\prime}%
}^{\prime}$ if it is possible for you, Alice and Bob, to use your own local
reversible operations (either individually or collectively) to transform
$\mathcal{Q}_{AB}$ and $\mathcal{Q}_{A^{\prime}B^{\prime}}^{\prime}$ into one
another. \ If this is possible, then $\mathcal{Q}_{AB}$ and $\mathcal{Q}%
_{A^{\prime}B^{\prime}}^{\prime}$ are of the \textbf{same entanglement type},
written
\[
\mathcal{Q}_{AB}\underset{loc}{\thicksim}\mathcal{Q}_{A^{\prime}B^{\prime}%
}^{\prime}%
\]

\noindent$\mathbb{Q}$\textsc{.}$\mathbb{E}$\textsc{. Yardstick 2}\textbf{.}
\ An EPR pair $\mathcal{Q}_{AB}$ \textbf{possesses more quantum entanglement
than} the NI$_{\mathbb{QE}}$ST standard EPR pair $\mathcal{Q}_{A^{\prime
}B^{\prime}}^{\prime}$ if it is possible for you, Alice and Bob, (either
individually or collectively)\ to apply your own reversible and irreversible
operations to your respective qubits to transform $\mathcal{Q}_{AB}$ into
$\mathcal{Q}_{A^{\prime}B^{\prime}}^{\prime}$. \ In this case, we write
\[
\mathcal{Q}_{AB}\underset{loc}{\geq}\mathcal{Q}_{A^{\prime}B^{\prime}}%
^{\prime}%
\]
\end{quotation}

\bigskip

\begin{quotation}
\noindent\textsc{Caveat}\textbf{.} \ Quantum entanglement may be irrevocably
lost if Quantum Entanglement Yardstick 2 is applied.

\bigskip
\end{quotation}

In summary, the above story about Alice and Bob has raised the following questions:

\begin{itemize}
\item \textbf{Question:} What type of entanglement do Alice and Bob
collectively possess?

\item \textbf{Question:} Is the quantum entanglement of $\mathcal{Q}_{AB}$ the
same as the quantum entanglement of $\mathcal{Q}_{A^{\prime}B^{\prime}%
}^{\prime}$ ?

\item \textbf{Question:} Is the quantum entanglement of $\mathcal{Q}_{AB}$
greater than the quantum entanglement of $\mathcal{Q}_{A^{\prime}B^{\prime}%
}^{\prime}$ ?
\end{itemize}

\bigskip

\section{Lest we forget, quantum entanglement is ...}

\bigskip

Before we continue with our story of Alice and Bob, now is a good opportunity
to restate the definition of quantum entanglement found in \cite{Lomonaco1}.
\ Readers not familiar with this definition or related concepts should refer
to sections 5 and 7 of \cite{Lomonaco1}.\bigskip

\begin{definition}
Let $\mathcal{Q}_{1}$, $\mathcal{Q}_{2}$, $\ldots$ , $\mathcal{Q}_{n}$ be
quantum systems with underlying Hilbert spaces $\mathcal{H}_{1}$,
$\mathcal{H}_{2}$, $\ldots$ , $\mathcal{H}_{n}$, respectively. \ And let
$\mathcal{Q}$ denote the global quantum system consisting of all the quantum
systems $\mathcal{Q}_{1}$, $\mathcal{Q}_{2}$, $\ldots$ , $\mathcal{Q}_{n}$,
where $\mathcal{H}=\bigotimes_{j=1}^{n}\mathcal{H}_{j}$ denotes the underlying
Hilbert space of $\mathcal{Q}$. \ Finally let the density operator $\rho$ on
the Hilbert space $\mathcal{H}$ denote the state of the global quantum system
$\mathcal{Q}$. \ Then $\mathcal{Q}$ is said to be \textbf{entangled} with
respect to the Hilbert space decomposition
\[
\mathcal{H}=%
{\displaystyle\bigotimes\limits_{j=1}^{n}}
\mathcal{H}_{j}%
\]
if it can not be written in the form
\[
\rho=\sum_{k=1}^{K}\lambda_{k}\left(  \bigotimes\limits_{j=1}^{n}\rho
_{(j,k)}\right)  \text{ ,}%
\]
for some positive integer $K$, where the $\lambda_{k}$'s are positive real
numbers such that
\[
\sum_{k=1}^{K}\lambda_{k}=1\text{ ,}%
\]
and where each $\rho_{(j,k)}$ is a density operator on the Hilbert space
$\mathcal{H}_{j}$. \ If $\rho$ is a pure state, then $\mathcal{Q}$ is
\textbf{entangled} if $\rho$ can not be written in the form
\[
\rho=\bigotimes\limits_{j=1}^{n}\rho_{j}\text{ ,}%
\]
where $\rho_{j}$ is a density operator on the Hilbert space $\mathcal{H}_{j}$.
\end{definition}

\bigskip

\section{Back to Alice and Bob: Local Moves and the Fundamental Problem of
Quantum Entanglement (FPQE)}

\bigskip

Although the story of Alice and Bob was told with two qubits, the same story
could have been told instead with three people, Alice, Bob, Cathy, and three
qubits. \ Or for that matter, it could have equally been told for $n$ people
with $n$ qubits. \ From now on, we will consider the more general story of $n$
people and $n$ qubits.

\bigskip

What Alice, Bob, Cathy, et al were trying to understand can be stated most
succinctly as the Fundamental Problem of quantum entanglement, namely:

\bigskip

\noindent\textbf{Fundamental Problem of Quantum Entanglement} (\textbf{FPQE}).
Let $\rho$ and $\rho^{\prime}$ be density operators representing two different
states of a quantum system $\mathcal{Q}$. \ Is it possible to move
$\mathcal{Q}$ from state $\rho$ to state $\rho^{\prime}$ by applying only
\textbf{local moves}?

\bigskip%

\[
\fbox{$%
\begin{array}
[c]{c}%
\text{But what is meant by the phrase ``\textbf{local move'' }?}%
\end{array}
$}%
\]

\bigskip

We define the \textbf{standard local moves} as:

\bigskip

\begin{definition}
The \textbf{standard local moves} are:
\end{definition}

\begin{itemize}
\item Local unitary transformations of the form
\[%
{\displaystyle\bigotimes\limits_{k=1}^{n}}
U_{k}\in%
{\displaystyle\bigotimes\limits_{k=1}^{n}}
\mathbb{U}\left(  \mathcal{H}_{k}\right)
\]
For example, for bipartite quantum systems, unitary transformations of the
form $U_{A}\otimes I$, $I\otimes U_{B}$, $U_{A}\otimes U_{B}$

\item Measurement of local observables of the form
\[%
{\displaystyle\bigotimes\limits_{k=1}^{n}}
\mathcal{O}_{k}\in%
{\displaystyle\bigotimes\limits_{k=1}^{n}}
Observables\left(  \mathcal{H}_{k}\right)
\]
\end{itemize}

\bigskip

\begin{example}
For example, for bipartite quantum systems\footnote{A bipartite quantum system
is a global quantum system consisting of two quantum systems.}, measurement of
local observables of the form $\mathcal{O}_{A}\otimes I$, $I\otimes
\mathcal{O}_{B}$, $\mathcal{O}_{A}\otimes\mathcal{O}_{B}$
\end{example}

\bigskip

We also define the \textbf{extended local moves} as

\bigskip

\begin{definition}
The \textbf{extended local moves} are:
\end{definition}

\begin{itemize}
\item Extended local unitary transformations of the form
\[%
{\displaystyle\bigotimes\limits_{k=1}^{n}}
\mathbb{U}\left(  \mathcal{H}_{k}\otimes\widetilde{\mathcal{H}}_{k}\right)
\text{ ,}%
\]
where $\mathcal{H}_{1}$, $\widetilde{\mathcal{H}}_{1}$, $\ldots$ ,
$\mathcal{H}_{n}$, $\widetilde{\mathcal{H}}_{n}$ are distinct non-overhapping
Hilbert spaces

\item Measurement of extended local observables of the form
\[%
{\displaystyle\bigotimes\limits_{k=1}^{n}}
Observables\left(  \mathcal{H}_{k}\otimes\widetilde{\mathcal{H}}_{k}\right)
\text{ ,}%
\]
where $\mathcal{H}_{1}$, $\widetilde{\mathcal{H}}_{1}$, $\ldots$ ,
$\mathcal{H}_{n}$, $\widetilde{\mathcal{H}}_{n}$ are distinct non-overlapping
Hilbert spaces
\end{itemize}

\bigskip

\begin{definition}
Moves based on unitary transformation are called \textbf{reversible}. \ Those
based on measurement are called \textbf{irreversible.}
\end{definition}

\bigskip

The Horodecki's\cite{Horodecki1}, \cite{Horodecki2}, \cite{Horodecki3},
Jonathan\cite{Jonathan1}, \cite{Jonathan2}, Linden\cite{Linden1},
\cite{Linden2}, Nielsen\cite{Nielsen1}, \cite{Nielsen2}, \cite{Nielsen3},
\cite{Nielsen4}, Plenio\cite{Jonathan1}, \cite{Jonathan2},
Popescu\cite{Linden1}, \cite{Linden2}, \cite{Virmani1}, \cite{Lo1} have made
some progress in understanding the FPQE in terms of all four of the above
local moves. \ For the rest of the talk, we restrict our discussion to
reversible standard local moves.

\bigskip

\section{A momentary digression: \ Two different perspectives}

\bigskip

Before continuing, it should be mentioned that physics and mathematics
approach quantum mechanics from two slightly different but equivalent
viewpoints. \ To avoid possible confusion, we describe below the minor
terminology differences that arise from these two slightly different perspectives.

\bigskip

Physics describes the state of a quantum system in terms of a traceless
Hermitian operator $\rho$, called the density operator. \ \ Observables are
Hermitian operators $\mathcal{O}$. \ Quantum states change via unitary
transformations $U$ according to the rubric
\[
\rho\longmapsto U\rho U^{\dagger}\text{ \ .}%
\]

\bigskip

On the other hand, mathematics describes the state of a quantum system in
terms of a skew Hermitian operator $i\rho$, also called the density operator.
\ Observables are skew Hermitian operators $i\mathcal{O}$. \ Quantum dynamics
are defined via the rule
\[
i\rho\longmapsto Ad_{U}\left(  i\rho\right)  \text{ ,}%
\]
where $U$ is a unitary operator lying in the Lie group of unitary
transformations $\mathbb{U}\left(  N\right)  $, and where $Ad$ denotes the big
adjoint operator. \ Please note that both density operators $i\rho$ and the
observables $i\mathcal{O}$ lie in the Lie algebra $u\left(  N\right)  $ of the
unitary group $\mathbb{U}\left(  N\right)  $.

\bigskip

These minor, but nonetheless annoying differences are summarized in the table below.

\bigskip

$%
\begin{tabular}
[c]{||c||}\hline\hline%
\begin{tabular}
[c]{l||l}%
\textbf{Physics}\hspace{1.5in} & \hspace{1.5in}\textbf{Math}%
\end{tabular}
\\\hline\hline
\fbox{%
\begin{tabular}
[c]{c}%
Hilbert Space $\mathcal{H}$\\
$Dim(\mathcal{H})=N$\\
\multicolumn{1}{l}{%
\begin{tabular}
[c]{c}%
Unitary Group\\
Lie Group
\end{tabular}
$\mathbb{U}(N)$}%
\end{tabular}
}\\\hline\hline%
\begin{tabular}
[c]{||l||l||}\hline\hline
\quad\quad\quad%
\begin{tabular}
[c]{l}%
$\left.
\begin{tabular}
[c]{ll}%
Observables: & $\mathcal{O}$\\
Density Ops: & $\rho$%
\end{tabular}
\right\}  $\\
\multicolumn{1}{c}{%
\begin{tabular}
[c]{l}%
$N\times N$ Hermitian Ops\\
\multicolumn{1}{c}{$A^{\dagger}=\overline{A}^{T}=A$}%
\end{tabular}
}%
\end{tabular}
\quad\quad\quad &
\begin{tabular}
[c]{l}%
$\left.
\begin{tabular}
[c]{ll}%
Observables: & $i\mathcal{O}$\\
Density Ops: & $i\rho$%
\end{tabular}
\right\}  $\\%
\begin{tabular}
[c]{c}%
$N\times N$ skew Hermitian Ops $\in u(N)$\\
$\left(  iA\right)  ^{\dagger}=\overline{\left(  iA\right)  }^{T}=-iA$\\
\multicolumn{1}{l}{where $u(N)=$ Lie algebra of $\mathbb{U}(N)$}%
\end{tabular}
\end{tabular}
\\\hline\hline
\multicolumn{1}{||c||}{%
\begin{tabular}
[c]{l}%
Dynamics via $U\in\mathbb{U}(N)$\\
\\
\multicolumn{1}{c}{$\left|  \psi\right\rangle \longmapsto U\left|
\psi\right\rangle $}\\
\multicolumn{1}{c}{$\rho\longmapsto U\rho U^{\dagger}$}%
\end{tabular}
} & \multicolumn{1}{||c||}{%
\begin{tabular}
[c]{c}%
Dynamics via $U\in\mathbb{U}(N)$\\
\\
$\left|  \psi\right\rangle \longmapsto U\left|  \psi\right\rangle $\\
$i\rho\longmapsto Ad_{U}\left(  i\rho\right)  $\\
where $Ad_{U}(i\rho)=U(i\rho)U^{-1}$\\
\multicolumn{1}{l}{is the Big adjoint rep.}%
\end{tabular}
}\\\hline\hline
\end{tabular}
\\\hline\hline
\end{tabular}
$

\bigskip

We will use the two different terminologies and conventions interchangeably.
\ Which terminology we are using should be clear from context.

\bigskip

\begin{remark}
From \label{physical density operator}\cite{Lomonaco1} we know that an element
$i\rho$ of the Lie algebra $\mathbf{u}\left(  N\right)  $ is a physical
density operator if and only if $\rho$ is positive semi-definite and of trace
1. \ Thus, the set
\[
\mathbf{density}\left(  N\right)  =\left\{  i\rho\in\mathbf{u}\left(
N\right)  \mid\rho\text{ is positive semi-definite of trace 1}\right\}
\]
of physical density operators is a convex subset of the Lie algebra
$\mathbf{u}\left(  N\right)  $.
\end{remark}

\bigskip

\section{The Group of Local Unitary Transformations and the Restricted FPQE}

\bigskip

For the sake of clarity of exposition and for the purpose of avoiding minor
technicalities, from on we consider only qubit quantum systems, i.e., quantum
systems consisting of qubits. The reader, if he/she so wishes, should be able
to easily rephrase the results of this paper to more general quantum systems.

\bigskip

Moreover, from this point on, we limit the scope of this talk to the study of
quantum entanglement from the perspective of the standard local unitary
transformations, i.e., from the perspective of standard reversible local moves
as defined in section 5 of this paper. \ To emphasize this point, we define
the group of local unitary transformations $\mathbb{L}(2^{n})$ as follows:

\bigskip

\begin{definition}
The \textbf{group of local unitary transformations} $\mathbb{L}(2^{n})$ is the
subgroup of $\mathbb{U}(2^{n})$ defined by
\[
\mathbb{L}(2^{n})=\bigotimes_{1}^{n}\mathbb{SU}(2)\text{ ,}%
\]
where $\mathbb{SU}(2)$ denotes the special unitary group.
\end{definition}

\bigskip

Henceforth, the phrase \textbf{``local move''} will mean an element of the
group $\mathbb{L}(2^{n})$ of \textbf{local unitary transformations}..

\bigskip

\noindent\textbf{Convention. \ }From this point on,\textbf{ }
\[
\fbox{$%
\begin{array}
[c]{c}%
\text{\textbf{Local Moves }}=\mathbb{L}(2^{n})
\end{array}
$}%
\]

\bigskip

Thus, for the rest of this paper we consider only the \textbf{Restricted
Fundamental Problem of Quantum Entanglement} (\textbf{RFPQE}), which is
defined as follows:

\bigskip

\noindent\textbf{Restricted Fundamental Problem of Quantum Entanglement}
(\textbf{RFPQE}). \textit{Let }$i\rho$\textit{\ and }$i\rho^{\prime}%
$\textit{\ be density operators lying in the Lie algebra }$\mathbf{u}\left(
2^{n}\right)  $.\textit{\ \ Does there exist a local move }$U$, i.e., a
$U\in\mathbb{L}(2^{n})$ such that
\[
i\rho^{\prime}=U\left(  i\rho\right)  U^{\dagger}=Ad_{U}\left(  i\rho\right)
\text{ ?}%
\]

\bigskip

We will need the following definition:

\bigskip

\begin{definition}
Two elements $i\rho$ and $i\rho^{\prime}$ in $u(2^{n})$ are said to be
\textbf{locally equivalent (or, of the same entanglement type}), written
\[
i\rho\underset{loc}{\thicksim}i\rho^{\prime}%
\]
provided there exists a $U\in\mathbb{L}(2^{n})$ such that
\[
i\rho^{\prime}=Ad_{U}\left(  i\rho\right)  =U\left(  i\rho\right)  U^{-1}%
\]
The equivalence class
\[
\left[  i\rho\right]  _{E}=\left\{  i\rho^{\prime}\mid i\rho\underset
{loc}{\thicksim}i\rho^{\prime}\right\}
\]
is called an \textbf{entanglement class} (or, an \textbf{orbit} of the big
adjoint action of $\mathbb{L}(2^{n})$ on the Lie algebra $\mathbf{u}\left(
2^{n}\right)  $). \ Finally, let
\[
\mathbf{u}(2^{n})/\mathbb{L}(2^{n})
\]
denote the \textbf{set of entanglement classes}.
\end{definition}

\bigskip%

\begin{center}
\fbox{\includegraphics[
height=1.5947in,
width=3.8138in
]%
{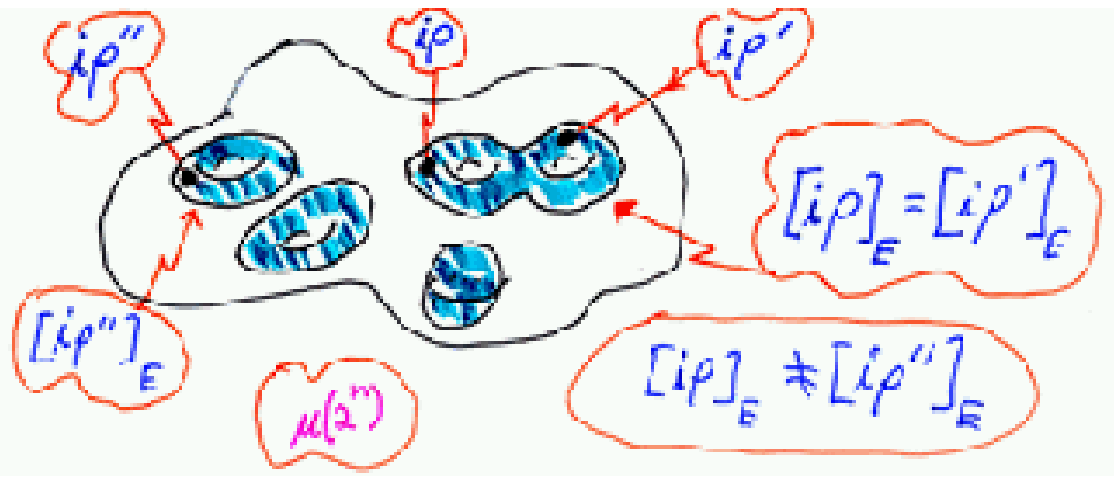}%
}\\
\textbf{Figure 2. Quantum entanglement classes.}%
\end{center}

\bigskip

The entanglement classes of the Lie algebra $\mathbf{u}(2^{n})$ are just the
\textbf{orbits} of the big adjoint action of $\mathbb{L}(2^{n})$ on
$\mathbf{u}(2^{n})$. \ Two states are entangled in the same way if and only if
they lie in the same entanglement class, i.e., in the same orbit.

\bigskip

\begin{remark}
Local unitary transformations can not entangle quantum systems with respect to
the above tensor product decomposition. \ However, \textbf{global unitary
transformations} \ (i.e., unitary transformations lying in $\mathbb{U}\left(
2^{n}\right)  -\mathbb{L}\left(  2^{n}\right)  $ ) are those unitary
transformations which can and often do produce interactions which entangle
quantum systems.
\end{remark}

\bigskip%

\[
\fbox{$%
\begin{array}
[c]{c}%
\text{\textbf{But what is quantum entanglement?}}%
\end{array}
$}%
\]

\bigskip

\section{Summary and List of Objectives}

\bigskip

We are now in a position to state clearly the main objectives of this paper.
\ Namely, in regard to the \textbf{Restricted Fundamental Problem of Quantum
Entanglement (RFPQE)}, our objectives are twofold:

\bigskip

\begin{itemize}
\item[\textbf{Objective 1.}] Given a density operator $i\rho$, devise a means
of determining the dimension of its entanglement class $\left[  i\rho\right]
_{E}$. \ We will accomplish this by determining the dimension of the tangent
plane $T_{i\rho}\left[  i\rho\right]  _{E}$ to the manifold $\left[
i\rho\right]  _{E}$ at the point $i\rho$. \ 
\end{itemize}

\bigskip

\begin{itemize}
\item[\textbf{Objective 2.}] Given two states $i\rho$ and $i\rho^{\prime}$,
devise a means of determining whether they belong to the same or different
entanglement class. \ We will accomplish this by constructing a
\textbf{complete set of quantum entanglement invariants}, i.e., invariants
that completely specify all the orbits ( i.e., all the entanglement classes).
\ In this sense, we have completely quantified quantum entanglement.\ \ In
other words, find a finite set $\left\{  f_{1},f_{2},\ldots,f_{K}\right\}  $
of real valued functions on $\mathbf{u}(2^{n})$ which distinguish all
entanglement classes, i.e.,
\[
i\rho\underset{loc}{\thicksim}i\rho^{\prime}\Longleftrightarrow f_{k}%
(i\rho)=f_{k}(i\rho^{\prime})\text{ for every }k\text{. }%
\]
\end{itemize}

\bigskip

\section{If you are unfamiliar with ... , then make a quantum jump to
Appendices A \& B}

\bigskip

This section is meant to play the role of a litmus test for the reader. \ If
the reader feels reasonably comfortable with the concepts listed below, then
it is suggested that the reader proceed to the next section of this paper.
\ If not, it is strongly suggested that the reader read Appendices A and B of
this paper before proceeding to the next section.

\bigskip

Let $\mathbb{G}$ be a Lie group, and let $\mathfrak{g}$ denote its Lie algebra.

\bigskip

\subsection{Litmus Test 1. \ \ The exponential map}

\qquad\bigskip

The reader should be familiar with the exponential map
\[
\exp:\mathfrak{g}\longrightarrow\mathbb{G}\text{ ,}%
\]
which for matrix Lie Groups is given by the power series
\[
\exp\left(  M\right)  =\sum_{k=0}^{\infty}\frac{1}{k!}M^{k}%
\]

\subsection{Litmus Test 2. \ The Lie bracket}

\qquad\bigskip

The reader should be familiar with the Lie bracket
\[
\left[  -,-\right]  :\mathfrak{g}\times\mathfrak{g}\longrightarrow
\mathfrak{g}\text{ ,}%
\]
which for matrix Lie groups is given by the commutator
\[
\left[  A,B\right]  =AB-BA
\]

\bigskip

\subsection{Litmus Test 3. \ The Lie algebra under three different guises}

\qquad\bigskip

The Lie algebra $\mathfrak{g}$ of the Lie group $\mathbb{G}$ can be viewed in
each of the following mathematically equivalent ways:

\begin{itemize}
\item As $T_{I}\mathbb{G}$, i.e., as the tangent space to the Lie group
$\mathbb{G}$ at the identity $I$.

\item As $\mathbf{Vec}_{R}\left(  \mathbb{G}\right)  $, i.e., as the Lie
algebra of right invariant smooth vector fields on the Lie group $\mathbb{G}$.

\item As $\mathbf{Der}_{\infty}\left(  \mathbb{G}\right)  $, i.e., as the Lie
algebra of all derivations (i.e., directional derivatives) on the algebra
$C^{\infty}\left(  \mathbb{G}\right)  $ of all smooth real valued functions on
$\mathbb{G}$.
\end{itemize}

\bigskip

In summary,
\[
\fbox{$\mathfrak{g}=T_{I}\mathbb{G}=\mathbf{Vec}_{R}\left(  \mathbb{G}\right)
=\mathbf{Der}_{\infty}\left(  \mathbb{G}\right)  $}%
\]

If you feel comfortable with the above three litmus tests, then please proceed
to the next section.

\bigskip

\section{Definition of Quantum Entanglement Invariants}

\bigskip

Let $C^{\infty}\left(  \mathbf{u}\left(  2^{n}\right)  \right)  $ denote the
algebra of smooth ($C^{\infty}$) real valued functions on the Lie algebra
$\mathbf{u}\left(  2^{n}\right)  $, i.e.,
\[
C^{\infty}\left(  \mathbf{u}\left(  2^{n}\right)  \right)  =\left\{
f:\mathbf{u}\left(  2^{n}\right)  \longrightarrow\mathbb{R}\mid f\text{ is
smooth}\right\}
\]

\bigskip

\begin{definition}
A function $f\in C^{\infty}\left(  \mathbf{u}\left(  2^{n}\right)  \right)  $
is called a \textbf{(quantum) entanglement invariant} if $f$ is invariant
under the big adjoint action of $\mathbb{L}\left(  2^{n}\right)  $, i.e., if
\[
f\left(  Ad_{U}(i\rho)\right)  =f(i\rho)
\]
for all $U\in\mathbb{L}\left(  2^{n}\right)  $, and for all $i\rho$ in
$\mathbf{u}\left(  2^{n}\right)  $. \ The collection of all (quantum)
entanglement invariants forms an algebra, which we denote by
\[
C^{\infty}\left(  \mathbf{u}\left(  2^{n}\right)  \right)  ^{\mathbb{L}\left(
2^{n}\right)  }\text{ .}%
\]
\end{definition}

\begin{definition}
A subset $\left\{  f_{1},f_{2},\ldots,f_{m}\right\}  $ of $C^{\infty}\left(
\mathbf{u}\left(  2^{n}\right)  \right)  ^{\mathbb{L}\left(  2^{n}\right)  }$
is called a \textbf{complete set of entanglement invariants} if
\[
i\rho\underset{loc}{\thicksim}i\rho^{\prime}\text{ iff }f_{k}\left(
i\rho\right)  =f_{k}\left(  i\rho^{\prime}\right)  \text{ for all }f_{k}\text{
in }\left\{  f_{1},f_{2},\ldots,f_{m}\right\}  \text{ .}%
\]
\end{definition}

\begin{definition}
Let $\mathcal{P}\left(  \mathbf{u}(2^{n})\right)  $ be the subalgebra of
$C^{\infty}\left(  \mathbf{u}\left(  2^{n}\right)  \right)  $ of all functions
$f\in C^{\infty}\left(  \mathbf{u}\left(  2^{n}\right)  \right)  $ which are
polynomial functions, i.e., of all functions $f$ for which $f\left(  v\right)
$ is a polynomial function of the entries in $v$. \ We define the
\textbf{algebra of polynomial entanglement invariants as }
\[
\mathcal{P}\left(  \mathbf{u}(2^{n})\right)  ^{\mathbb{L}\left(  2^{n}\right)
}=\mathcal{P}\left(  \mathbf{u}(2^{n})\right)  \cap C^{\infty}\left(
\mathbf{u}\left(  2^{n}\right)  \right)  ^{\mathbb{L}\left(  2^{n}\right)  }%
\]
\end{definition}

\bigskip

\begin{theorem}
$\mathcal{P}\left(  \mathbf{u}(2^{n})\right)  ^{\mathbb{L}\left(
2^{n}\right)  } $ is a finitely generated algebra.
\end{theorem}

\bigskip

\begin{definition}
A minimal set of generators of $\mathcal{P}\left(  \mathbf{u}(2^{n})\right)
^{\mathbb{L}\left(  2^{n}\right)  }$ is called a \textbf{basic set of
entanglement invariants}. \ 
\end{definition}

\bigskip

\begin{remark}
It is important to note that a basic set of entanglement invariants is not
always a complete set of entanglement invariants. \ We will show that this is
the case for entanglement invariants for two qubit quantum systems.
\end{remark}

\bigskip

\section{The Lie Algebra $\mathbf{\ell}\left(  2^{n}\right)  $ of
$\mathbb{L}(2^{n})$}

\bigskip

To understand and work with the big adjoint action \label{Section Lie Alg}
\[
\mathbb{L}\left(  2^{n}\right)  \times\mathbf{u}\left(  2^{n}\right)
\overset{Ad}{\longrightarrow}\mathbf{u}\left(  2^{n}\right)
\]
we will need to lift this action to the corresponding infinitesimal action of
the Lie algebra $\mathbf{\ell}\left(  2^{n}\right)  $ of $\mathbb{L}(n)$.
\ \ The Lie algebra $\mathbf{\ell}\left(  2^{n}\right)  $ will play a crucial
role in our achieving objectives 1 and 2 as stated in the previous section.

\bigskip

\begin{definition}
The Lie algebra $\mathbf{\ell}\left(  2^{n}\right)  $ is the (real) Lie
algebra given by the following Kronecker sum
\[
\mathbf{\ell}\left(  2^{n}\right)  =\underset{n\text{ terms}}{\underbrace
{\mathbf{su}\left(  2\right)  \boxplus\mathbf{su}\left(  2\right)
\boxplus\cdots\boxplus\mathbf{su}\left(  2\right)  }}\text{ ,}%
\]
where $\mathbf{su}\left(  2\right)  $ denotes the Lie algebra of the special
unitary group $\mathbb{SU}\left(  2\right)  $, and where the \textbf{Kronecker
sum} `$A\boxplus B$' of two matrices (or operators) $A$ and $B$ is defined by
\[
A\boxplus B=A\otimes\mathbf{1}+\mathbf{1}\otimes B\text{ ,}%
\]
with `$\mathbf{1}$' denoting the identity matrix (or operator).
\end{definition}

\bigskip

A basis\footnote{For more information, please refer to Appendix B.} of the
(real) Lie algebra $\mathbf{u}\left(  2^{n}\right)  $ is given by
\[
\left\{  \xi_{k_{1}k_{2}\ldots k_{n}}\mid k_{1},k_{2},\ldots,k_{n}%
=0,1,2,3\right\}  \text{ ,}%
\]
where
\[
\xi_{k_{1}k_{2}\ldots k_{n}}=-\frac{i}{2}\sigma_{k_{1}}\otimes\sigma_{k_{2}%
}\otimes\cdots\otimes\sigma_{k_{n}}\text{ ,}%
\]
and where
\[
\sigma_{1}=\left(
\begin{array}
[c]{cc}%
0 & 1\\
1 & 0
\end{array}
\right)  \text{, }\sigma_{2}=\left(
\begin{array}
[c]{cr}%
0 & -i\\
i & 0
\end{array}
\right)  \text{, }\sigma_{3}=\left(
\begin{array}
[c]{cr}%
1 & 0\\
0 & -1
\end{array}
\right)  \text{, }%
\]
denote the Pauli spin matrices, and where
\[
\sigma_{0}=\left(
\begin{array}
[c]{cc}%
1 & 0\\
0 & 1
\end{array}
\right)
\]
denotes the $2\times2$ identity matrix.

\bigskip

It follows that a basis of the Lie algebra $\mathbf{\ell}\left(  2^{n}\right)
$ as a subalgebra of $\mathbf{u}\left(  2^{n}\right)  $ is:
\[
\left\{  \xi_{k_{1}k_{2}\ldots k_{n}}\mid k_{1},k_{2},\ldots,k_{n}%
=0,1,2,3\text{, where exactly one }k_{j}\neq0\right\}
\]

\bigskip

For example,

\begin{itemize}
\item $\left\{  \xi_{1}\text{, }\xi_{2}\text{, }\xi_{3}\right\}  $ is a basis
of $\mathbf{\ell}\left(  1\right)  $

\item $\left\{  \xi_{01}\text{, }\xi_{02}\text{, }\xi_{03}\text{, }\xi
_{10}\text{, }\xi_{20}\text{, }\xi_{30}\right\}  $ is a basis of
$\mathbf{\ell}\left(  2\right)  $

\item $\left\{  \xi_{001}\text{, }\xi_{002}\text{, }\xi_{003}\text{, }%
\xi_{010}\text{, }\xi_{020}\text{, }\xi_{030}\text{, }\xi_{100}\text{, }%
\xi_{200}\text{, }\xi_{300}\right\}  $ is a basis of $\mathbf{\ell}\left(
3\right)  $
\end{itemize}

\bigskip

Thus, we have the following proposition:

\bigskip

\begin{proposition}
The Lie algebra $\mathbf{\ell}\left(  2^{n}\right)  $ of the Lie group
$\mathbb{L}\left(  2^{n}\right)  $ of local unitary transformations is of
dimension $3n$, i.e.,
\[
Dim\left(  \ \mathbf{\ell}\left(  2^{n}\right)  \ \right)  =3n
\]
\end{proposition}

\bigskip

\section{Definition of the Infinitesimal Action}

\bigskip

We now show how quantum entanglement invariants can be found by lifting the
big adjoint action to the Lie algebra $\mathbf{\ell}\left(  2^{n}\right)  $,
where the problem becomes a linear one. \label{Section Act Def}

\bigskip

The big adjoint action
\[
\mathbb{L}\left(  2^{n}\right)  \times\mathbf{u}\left(  2^{n}\right)
\overset{Ad}{\longrightarrow}\mathbf{u}\left(  2^{n}\right)
\]
induces an \textbf{infinitesimal} \textbf{action}
\[
\mathbf{\ell}\left(  2^{n}\right)  \overset{\Omega}{\longrightarrow
}\mathbf{Vec}\left(  \mathbf{u}(2^{n})\right)
\]
as follows.

\bigskip

Let $v\in\mathbf{\ell}\left(  2^{n}\right)  $. \ We define the vector field
$\Omega\left(  v\right)  $ on $\mathbf{u}\left(  2^{n}\right)  $ by
constructing a tangent vector $\left.  \Omega\left(  v\right)  \right|
_{i\rho}$ for each $i\rho\in\mathbf{u}\left(  2^{n}\right)  $.

\bigskip

Let $\gamma_{v}\left(  t\right)  $ be the smooth curve in $\mathbf{u}\left(
2^{n}\right)  $ defined by
\[
\gamma_{v}\left(  t\right)  =Ad_{\exp\left(  tv\right)  }\left(  i\rho\right)
\text{ .}%
\]
Then $\gamma_{v}\left(  t\right)  $ is a curve which passes through $i\rho$ at
time $t=0$. \ We define $\left.  \Omega\left(  v\right)  \right|  _{i\rho}$ as
the tangent vector to $\gamma_{v}\left(  t\right)  $ at $t=0$.%

\begin{center}
\fbox{\includegraphics[
height=1.7495in,
width=3.4264in
]%
{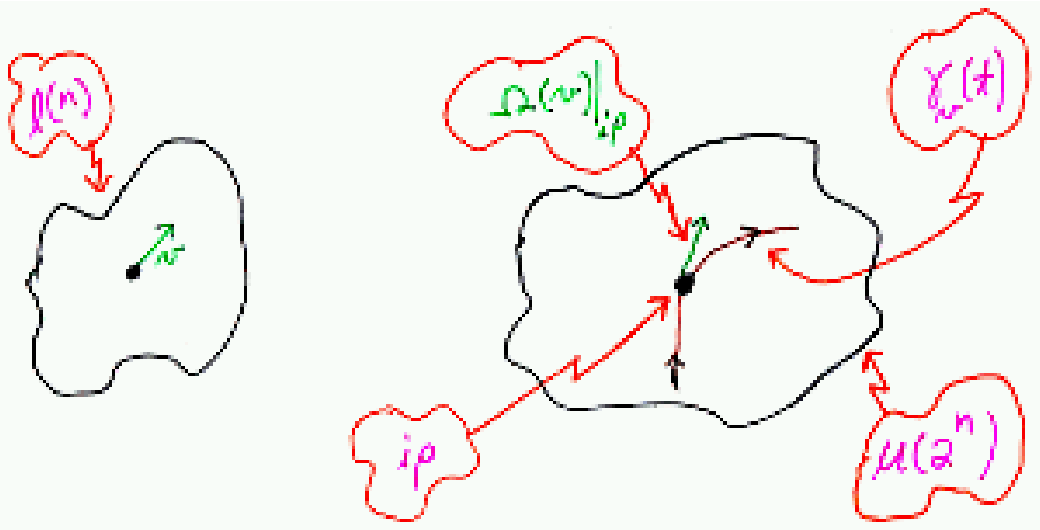}%
}\\
\textbf{Figure 3. \ Induced infinitesimal action.}%
\end{center}

\bigskip

But what is the meaning of the infinitesimal action
\[
\mathbf{\ell}\left(  2^{n}\right)  \overset{\Omega}{\longrightarrow
}\mathbf{Vec}\left(  \mathbf{u}(2^{n})\right)
\]
that we have just defined? \ 

\bigskip

Each $\left.  \Omega\left(  i\rho\right)  \right|  _{i\rho}$ is a direction in
$\mathbf{u}\left(  2^{n}\right)  $ from $i\rho$ that we can move without
leaving the quantum entanglement class $\left[  i\rho\right]  _{E}$.
\ Movement in all directions not in $\left.  \operatorname{Im}\left(
\Omega\right)  \right|  _{i\rho}$ will force us to immediately leave $\left[
i\rho\right]  _{E}$. \ %

\begin{center}
\fbox{\includegraphics[
height=1.8697in,
width=3.2456in
]%
{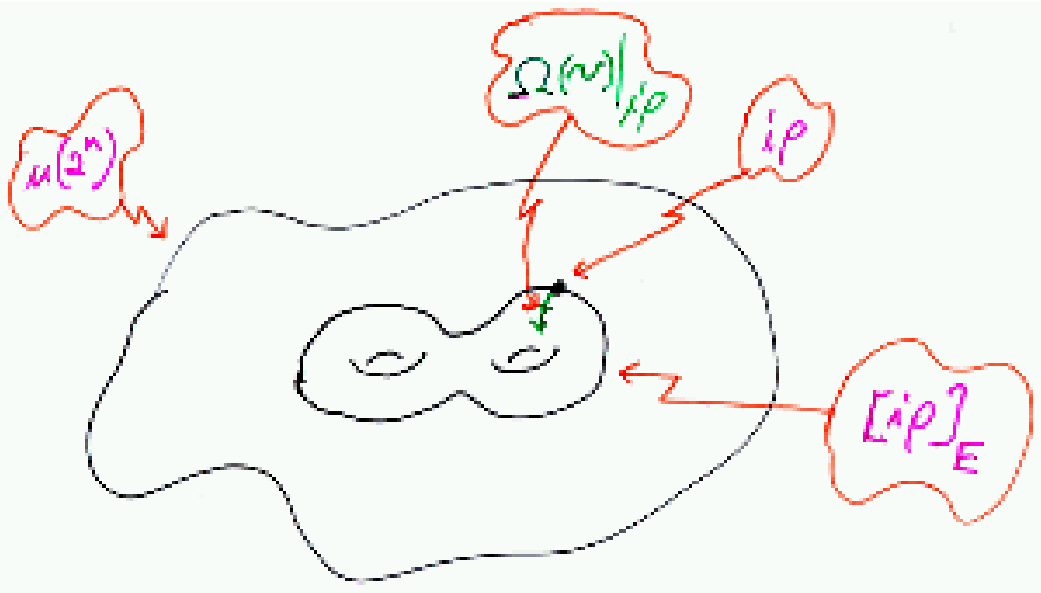}%
}\\
\textbf{Figure 4. \ Moving in a direction that stays within the entanglement
class.}%
\end{center}

\bigskip

As the reader might expect, the infinitesimal action $\Omega$ can naturally be
expressed in terms of the small adjoint operator $ad$. \ In particular, we have:

\bigskip

\begin{proposition}
$\Omega\left(  v\right)  =ad_{v}$ for all $v$ in the Lie algebra
$\mathbf{\ell}\left(  2^{n}\right)  $ \label{ad}
\end{proposition}

\begin{proof}
Let $v$ be an arbitrary element of the Lie algebra $\mathbf{\ell}\left(
2^{n}\right)  $, and let $i\rho$ be an arbitrary element of the Lie algebra
$\mathbf{u}\left(  2^{n}\right)  $.

By definition, $\Omega\left(  v\right)  (i\rho)$ is the tangent vector at
$t=0$ to the curve $\gamma_{v}\left(  t\right)  $ in $\mathbf{u}\left(
2^{n}\right)  $ given by
\[
\gamma_{v}\left(  t\right)  =Ad_{\exp\left(  tv\right)  }\left(  i\rho\right)
\text{ .}%
\]
Hence,
\[%
\begin{array}
[c]{ccl}%
\Omega\left(  v\right)  \left(  i\rho\right)  & = & \frac{d}{dt}\left.
Ad_{\exp\left(  tv\right)  }\left(  i\rho\right)  \right|  _{t=0}\\
&  & \\
& = & \frac{d}{dt}\left.  \exp\left(  ad_{tv}\right)  \left(  i\rho\right)
\right|  _{t=0}\\
&  & \\
& = & \frac{d}{dt}\left.  \exp\left(  t\cdot ad_{v}\right)  \left(
i\rho\right)  \right|  _{t=0}\\
&  & \\
& = & \frac{d}{dt}\left.  \left(  1+t\cdot ad_{v}+o(t^{2})\right)  \left(
i\rho\right)  \right|  _{t=0}\\
&  & \\
& = & ad_{v}\left(  i\rho\right)
\end{array}
\]
\end{proof}

\bigskip

The above formula will prove to be useful when we actually calculate the
entanglement invariants of some examples given in later sections.

\bigskip

\section{The significance of the infinitesimal action $\Omega$}

\bigskip

\bigskip

As stated in the appendices, the Lie algebra $\mathbf{Vec}\left(
\mathbf{u}\left(  2^{n}\right)  \right)  $ of all smooth vector fields on
$\mathbf{u}\left(  2^{n}\right)  $ can be identified with the Lie algebra of
derivations $\mathbf{Der}\left(  C^{\infty}\mathbf{u}\left(  2^{n}\right)
\right)  $.

\bigskip

The significance of the infinitesimal action
\[
\Omega:\mathbf{\ell}\left(  2^{n}\right)  \longrightarrow\mathbf{Vec}\left(
\mathbf{u}\left(  2^{n}\right)  \right)
\]
is best expressed in terms of the following theorem:

\bigskip

\begin{theorem}
Let \label{Theorem ad}
\[
\left\{  v_{1},v_{2},\ldots,v_{3n}\right\}
\]
be a basis for the Lie algebra $\mathbf{\ell}\left(  2^{n}\right)  $. \ Then a
smooth real valued function
\[
f:\mathbf{u}\left(  2^{n}\right)  \longrightarrow\mathbb{R}%
\]
is an entanglement invariant if and only if it satisfies the following system
of partial differential equations
\[
\left\{
\begin{array}
[c]{ccc}%
\Omega\left(  v_{1}\right)  f & = & 0\\
&  & \\
\Omega\left(  v_{2}\right)  f & = & 0\\
&  & \\
\vdots & \vdots & \vdots\\
&  & \\
\Omega\left(  v_{3n}\right)  f & = & 0
\end{array}
\right.
\]
\end{theorem}

\bigskip

The intuition underlying the above theorem is that $\Omega\left(
v_{1}\right)  $, $\Omega\left(  v_{2}\right)  $, $\ldots$ , $\Omega\left(
v_{3n}\right)  $ are the linearly independent directions we can move without
leaving the entanglement class we are currently in. \ Hence, if $f$ is an
entanglement invariant, then its rate of change (i.e., its directional
derivative) in each of the directions $\Omega\left(  v_{1}\right)  $,
$\Omega\left(  v_{2}\right)  $, $\ldots$ , $\Omega\left(  v_{3n}\right)  $
must be zero, and vice versa.

\bigskip

This theorem provides us with a means of determining a complete set of
entanglement invariants. \ All that we need to do is to solve the above system
of partial differential equations.

\bigskip

\section{Achieving our two objectives, ... finally}

\bigskip

We now show how the infinitesimal action
\[
\mathbf{\ell}\left(  2^{n}\right)  \overset{\Omega}{\longrightarrow
}\mathbf{Vec}\left(  \mathbf{u}(2^{n})\right)
\]
can be used to achieve the two objectives listed in section 8 of this paper.

\bigskip

\begin{itemize}
\item[\textbf{Objective 1.}] Given an arbitrary density operator $i\rho$,
devise a means of determining the dimension of its entanglement class $\left[
i\rho\right]  _{E}$. \ 
\end{itemize}

\bigskip

Objective 1 is achieved as follows:

\bigskip

We begin by noting that $\left.  \mathbf{Vec}\left(  \mathbf{u}\left(
2^{n}\right)  \right)  \right|  _{i\rho}$ is the same as the tangent space
$T_{i\rho}\left(  \mathbf{u}(2^{n})\right)  $ to $\mathbf{u}\left(
2^{n}\right)  $ at the point $i\rho$, and that $\left.  \operatorname{Im}%
\left(  \Omega\right)  \right|  _{i\rho}$ is the same as the tangent space
$T_{i\rho}\left(  \left[  i\rho\right]  _{E}\right)  $ to the entanglement
class $\left[  i\rho\right]  _{E}$ at $i\rho$. \ Hence, the dimension of
$\left[  i\rho\right]  _{E}$ is same as the dimension as its tangent space at
$i\rho$, i.e.,
\[
Dim\left(  \left[  i\rho\right]  _{E}\right)  =Dim\left(  T_{i\rho}\left(
\left[  i\rho\right]  _{E}\right)  \right)  =Dim\left(  \left.
\operatorname{Im}\left(  \Omega\right)  \right|  _{i\rho}\right)
\]

The task of finding the dimension of the entanglement class $\left[
i\rho\right]  _{E}$ reduces to that of computing the dimension of the vector
space $\left.  \operatorname{Im}\left(  \Omega\right)  \right|  _{i\rho}$.
\ We will give examples of this dimension calculation in the next two sections.

\bigskip

We next use the infinitesimal action to achieve:

\bigskip

\begin{itemize}
\item[\textbf{Objective 2.}] Given two states $i\rho$ and $i\rho^{\prime}$,
devise a means of determining whether they belong to the same or different
entanglement class. \ \bigskip
\end{itemize}

\noindent as follows:

\bigskip

We begin by noting that $\mathbf{Vec}\left(  \mathbf{u}\left(  2^{n}\right)
\right)  $ can be identified with the Lie algebra $Der\left(  C^{\infty
}\mathbf{u}(2^{n})\right)  $ of derivations on $\mathbf{u}\left(
2^{n}\right)  $. \ Next we recall that $\operatorname{Im}\left(
\Omega\right)  $ consists of all directions in $\mathbf{u}\left(
2^{n}\right)  $ that we can move without leaving an entanglement class that we
are in. \ If
\[
f\in\left(  C^{\infty}\left(  \mathbf{u}(2^{n})\right)  \right)
^{\mathbb{L}\left(  2^{n}\right)  }%
\]
is an entanglement invariant, then $f$ will not change if we move in any
direction within $\operatorname{Im}\left(  \Omega\right)  $. \ As a result we
have the following theorem:

\bigskip

\begin{theorem}
Let $v_{1},v_{2},\ldots,v_{3n}$ be a vector space basis of the (real) Lie
algebra $\mathbf{\ell}\left(  2^{n}\right)  $. \ Then \
\[
f\in\left(  C^{\infty}\left(  \mathbf{u}(2^{n})\right)  \right)
^{\mathbb{L}\left(  2^{n}\right)  }\Longleftrightarrow\Omega\left(
v_{j}\right)  f=0
\]
for all $j$, where $\Omega\left(  v_{j}\right)  $ is interpreted as a
differential operator in $Der\left(  C^{\infty}\mathbf{u}(2^{n})\right)  $.
\end{theorem}

\bigskip

In other words, the task of finding entanglement invariants reduces to that of
solving a system of linear partial differential equations. \ We will give
examples of this calculation in the examples found in the next two sections of
this paper.

\bigskip

\section{Example 1. The entanglement classes of $n=1$ qubits}

\bigskip

We now make use of the methods developed in the previous section to study the
entanglement classes associated with $n=1$ qubits. \ This is a trivial but
nonetheless instructive case. \ As we shall see, there is no entanglement in
this case. \ But there are many entanglement classes!

\bigskip

For this example, the local unitary group $\mathbb{L}\left(  2^{1}\right)  $
is the same as the special unitary group $\mathbb{SU}\left(  2^{1}\right)  $.
\ The corresponding Lie algebra $\mathbf{\ell}\left(  2^{1}\right)  $ is the
same as the Lie algebra $\mathbf{su}(2)$. \ Each density operator $i\rho$ lies
in the Lie algebra $\mathbf{u}\left(  2^{1}\right)  $. \ 

\bigskip

As an immediate consequence of Proposition \ref{ad} of Section \ref{Section
Act Def}, the infinitesimal action
\[
\Omega:\mathbf{\ell}\left(  2^{1}\right)  \longrightarrow\mathbf{Vec}\left(
\mathbf{u}\left(  2\right)  \right)
\]
is simply the small adjoint action, i.e.,
\[
\Omega\left(  v\right)  =ad_{v}\text{ ,}%
\]
for all $v\in\mathbf{\ell}\left(  2^{1}\right)  $.

\bigskip

We can now use the bases\footnote{See Section \ref{Section Lie Alg}.}
\[%
\begin{array}
[c]{c}%
\left\{  \xi_{1}=-\frac{1}{2}\sigma_{1},\ \xi_{2}=-\frac{1}{2}\sigma_{2}%
,\ \xi_{3}=-\frac{1}{2}\sigma_{3}\right\} \\
\text{ \ and \ }\\
\left\{  \xi_{0}=-\frac{1}{2}\sigma_{0},\ \xi_{1}=-\frac{1}{2}\sigma_{1}%
,\ \xi_{2}=-\frac{1}{2}\sigma_{2},\ \xi_{3}=-\frac{1}{2}\sigma_{3}\right\}
\end{array}
\]
of the respective Lie algebras $\mathbf{\ell}\left(  2^{1}\right)  $ and
$\mathbf{u}\left(  2\right)  $ to find a more useful expression for
$\Omega\left(  v\right)  $. \ 

\bigskip

Each element $v\in\mathbf{\ell}\left(  2^{1}\right)  $ can be uniquely
expressed in the form
\[
v=a\cdot\xi\text{ ,}%
\]
where $a=\left(  a_{1},a_{2},a_{3}\right)  \in\mathbb{R}^{3}$ and $\xi=\left(
\xi_{1},\xi_{2},\xi_{3}\right)  $. \ Thus,
\[
\Omega\left(  v\right)  =\Omega\left(  a\cdot\xi\right)  =ad_{a\cdot\xi
}=a\cdot ad_{\xi}\text{ ,}%
\]
where
\[
ad_{\xi}=\left(  ad_{\xi_{1}},ad_{\xi_{2}},ad_{\xi_{3}}\right)  \text{ .}%
\]

\bigskip

Moreover, each element $i\rho\in\mathbf{u}\left(  2\right)  $ can be uniquely
written in terms of the basis of $\mathbf{u}\left(  2\right)  $ as
\[
i\rho=x_{0}\xi_{0}+x\cdot\xi\text{ ,}%
\]
where $x=\left(  x_{1},x_{2},x_{3}\right)  $ and $\xi=\left(  \xi_{1},\xi
_{2},\xi_{3}\right)  $.

\bigskip

In terms of the basis of $\mathbf{u}\left(  2\right)  $,
\[
ad_{\xi_{j}}=\left\{
\begin{array}
[c]{ll}%
\left(
\begin{array}
[c]{cc}%
0 & 0\\
0 & L_{j}%
\end{array}
\right)  =0\oplus L_{j} & \text{if }j=1,2,3\\
& \\
\left(
\begin{array}
[c]{cc}%
0 & 0\\
0 & 0
\end{array}
\right)  =0 & \text{if }j=0
\end{array}
\right.  \text{ ,}%
\]
where
\[
L_{1}=\left(
\begin{array}
[c]{rrr}%
0 & 0 & 0\\
0 & 0 & -1\\
0 & 1 & 0
\end{array}
\right)  \text{, }L_{2}=\left(
\begin{array}
[c]{rrr}%
0 & 0 & 1\\
0 & 0 & 0\\
-1 & 0 & 0
\end{array}
\right)  \text{, }L_{3}=\left(
\begin{array}
[c]{rrr}%
0 & -1 & 0\\
1 & 0 & 0\\
0 & 0 & 0
\end{array}
\right)
\]
is the basis\footnote{This follows from the following calculation:
\[%
\begin{array}
[c]{cclllll}%
ad_{\xi_{j}}\left(  \xi_{k}\right)  & = & ad_{-i\sigma_{j}/2}\left(
-i\sigma_{k}/2\right)  & = & \left[  -i\sigma_{j}/2,-i\sigma_{k}/2\right]  &
= & -\frac{1}{4}\left[  \sigma_{j},\sigma_{k}\right] \\
& = & -\frac{1}{2}i\epsilon_{jkp}\sigma_{p} & = & \epsilon_{jkp}\xi_{p} &  &
\end{array}
\]
where $L_{j}=\left(  \epsilon_{jkp}\right)  $.} of the Lie algebra
$\mathbf{so}\left(  3\right)  $ of the special orthogonal group $\mathbb{SO}%
\left(  3\right)  $ given in Appendix B.

\bigskip

Let
\[
\left\{  \frac{\partial}{\partial x_{0}}\text{, }\frac{\partial}{\partial
x_{1}}\text{, }\frac{\partial}{\partial x_{2}}\text{, }\frac{\partial
}{\partial x_{3}}\right\}
\]
denote the basis\footnote{For those unfamiliar with this basis, please refer
to Appendix A page \pageref{vec basis}.} of $\mathbf{Vec}\left(
\mathbf{u}\left(  2\right)  \right)  $ induced by the chart
\[%
\begin{array}
[c]{ccc}%
\mathbf{u}\left(  2\right)  & \overset{\pi}{\longrightarrow} & \mathbb{R}%
^{4}\\
&  & \\
i\rho=\sum_{j=0}^{3}x_{j}\xi_{j} & \longmapsto & \left(  x_{0},x_{1}%
,x_{2},x_{3}\right)  =\left(  x_{0},x\right)
\end{array}
\text{ .}%
\]
In other words, for each $j$, $\partial/\partial x_{j}$ denotes the vector
field on $\mathbf{u}\left(  2\right)  $ defined at each point $i\rho$ as the
tangent vector to the curve $\pi^{-1}\left(  x_{0},\ldots,x_{j}+t,\ldots
,x_{3}\right)  =i\rho+t\xi_{j}$ at $t=0$.

\bigskip

Then,
\[%
\begin{array}
[c]{ccl}%
\Omega\left(  v\right)  \left(  i\rho\right)  & = & \left(  x_{0},x\right)
\cdot\left(  0\oplus a\cdot L\right)  \cdot\left(
\begin{array}
[c]{c}%
\partial/\partial x_{0}\\
\partial/\partial x_{1}\\
\partial/\partial x_{2}\\
\partial/\partial x_{3}%
\end{array}
\right) \\
&  & \\
& = & x\cdot\left(  a\cdot L\right)  \cdot\bigtriangledown\\
&  & \\
& = & a\cdot x\times\bigtriangledown\text{ ,}%
\end{array}
\text{ ,}%
\]
where `$\times$' denotes the vector cross product, and where
\[
L=\left(  L_{1},L_{2},L_{3}\right)  \text{ and }\bigtriangledown=\left(
\begin{array}
[c]{c}%
\partial/\partial x_{1}\\
\partial/\partial x_{2}\\
\partial/\partial x_{3}%
\end{array}
\right)  \text{ .}%
\]

\bigskip

We can now achieve objective 1.

\bigskip

\noindent\textbf{Objective 1.} \ \textit{Given an arbitrary density operator}
$i\rho$ \textit{in} $\mathbf{u}\left(  2\right)  $, \textit{find the dimension
of an arbitrary entanglement class} $\left[  i\rho\right]  _{E}$.

\bigskip

From the above discussion, it follows that the image $\operatorname{Im}\left(
\Omega\right)  $ of the infinitesimal action $\Omega$ is spanned by the three
vector fields
\[
\left\{
\begin{array}
[c]{ccc}%
\Omega\left(  \xi_{1}\right)  & = & x_{3}\frac{\partial}{\partial x_{2}}%
-x_{2}\frac{\partial}{\partial x_{3}}\\
&  & \\
\Omega\left(  \xi_{2}\right)  & = & x_{1}\frac{\partial}{\partial x_{3}}%
-x_{3}\frac{\partial}{\partial x_{1}}\\
&  & \\
\Omega\left(  \xi_{3}\right)  & = & x_{2}\frac{\partial}{\partial x_{1}}%
-x_{1}\frac{\partial}{\partial x_{2}}%
\end{array}
\right.  \text{ ,}%
\]
defined on $\mathbf{u}\left(  2^{n}\right)  =\left\{  i\rho=\sum_{j=0}%
^{3}x_{j}\xi\mid x_{0},x_{1},x_{2},x_{3}\in\mathbb{R}\right\}  $. \ In
particular, the tangent space $T_{i\rho}\left(  \left[  i\rho\right]
_{E}\right)  =\left.  \left(  \operatorname{Im}\Omega\right)  \right|
_{i\rho}$ of the entanglement class $\left[  i\rho\right]  _{E}$ at the point
$i\rho$ is spanned by
\[
\left.  \Omega\left(  \xi_{1}\right)  \right|  _{i\rho}\text{, }\left.
\Omega\left(  \xi_{2}\right)  \right|  _{i\rho}\text{, }\left.  \Omega\left(
\xi_{3}\right)  \right|  _{i\rho}%
\]
As can be easily verified by the reader, the above three vectors span a two
dimensional space if $\left|  x\right|  \neq0$ and a zero dimensional vector
space if $\left|  x\right|  =0$. \ 

\bigskip

Since $\left.  \left(  \operatorname{Im}\Omega\right)  \right|  _{i\rho}$ is
the tangent space $T_{i\rho}\left(  \left[  i\rho\right]  _{E}\right)  $ of
$\left[  i\rho\right]  _{E}$ at the point $i\rho$, and since the dimension of
$\left[  i\rho\right]  _{E}$ is the same as the dimension of its tangent space
$T_{i\rho}\left(  \left[  i\rho\right]  _{E}\right)  $ at $i\rho$, it follows
that the dimension of the entanglement class $\left[  i\rho\right]  _{E}$ is
given by:
\[
Dim\left[  i\rho\right]  _{E}=\left\{
\begin{array}
[c]{ccc}%
2 & \text{if} & \left|  x\right|  \neq0\\
&  & \\
0 & \text{if} & \left|  x\right|  \neq0
\end{array}
\right.
\]

\bigskip

We are now ready to achieve objective 2:

\bigskip

\noindent\textbf{Objective 2.} \ \textit{Given two states} $i\rho$
\textit{and} $i\rho^{\prime}$, \textit{devise a means of determining whether
they belong to the same or different entanglement class}. \ 

\bigskip

We achieve this objective by determining a complete set of entanglement
invariants\footnote{As we shall see, in this particular case of $n=1$ qubits,
the complete set of entanglement invariants consists of only one invariant.}
for one qubit quantum systems, i.e., by determining a set of entanglement
invariants $\left\{  f_{1},f_{2},\ldots,f_{k}\right\}  $ such that
\[
i\rho\underset{loc}{\thicksim}i\rho^{\prime}\text{ \ if and only if \ }%
f_{j}\left(  i\rho\right)  =f_{j}\left(  i\rho^{\prime}\right)  \text{ for all
}j\text{ .}%
\]
.

\bigskip

We begin by recalling that the Lie algebra $\mathbf{Vec}\left(  \mathbf{u}%
\left(  2\right)  \right)  $ of vector fields on $\mathbf{u}\left(  2\right)
$ can be identified with the Lie algebra $\mathbf{Der}\left(  C^{\infty
}\mathbf{u}(2)\right)  $ of all derivations on the smooth real valued
functions on $\mathbf{u}(2)$. \ Thus, the elements of $\operatorname{Im}%
\Omega$ can be viewed as directional derivatives, directional derivatives in
those directions in which we can move and still remain in the same
entanglement class. \ 

\bigskip

From theorem \ref{Theorem ad},\ it immediately follows that a real valued
function $f:\mathbf{u}(2)\longrightarrow\mathbb{R}$ is an entanglement
invariant if and only it is a solution of the system of partial differential
equations (PDEs): \
\[
\left\{
\begin{array}
[c]{c}%
\Omega\left(  \xi_{1}\right)  f=0\\
\\
\Omega\left(  \xi_{2}\right)  f=0\\
\\
\Omega\left(  \xi_{3}\right)  f=0
\end{array}
\right.
\]

Since from above we know that $\Omega\left(  \xi_{j}\right)  \left(
i\rho\right)  =x\cdot L_{j}\cdot\bigtriangledown$, we can write the above
system of PDEs more explicitly as:
\[
\left\{
\begin{array}
[c]{c}%
x_{3}\frac{\partial f}{\partial x_{2}}-x_{2}\frac{\partial f}{\partial x_{3}%
}=0\\
\\
x_{1}\frac{\partial f}{\partial x_{3}}-x_{3}\frac{\partial f}{\partial x_{1}%
}=0\\
\\
x_{2}\frac{\partial f}{\partial x_{1}}-x_{1}\frac{\partial f}{\partial x_{2}%
}=0
\end{array}
\right.  \text{ ,}%
\]
where, as before, $i\rho=x_{0}\xi_{0}+x\cdot\xi$.

\bigskip

From theorem \ref{Theorem ad}, we know that a complete set of quantum
entanglement invariants for one qubit systems is the same as a complete
functionally independent set of solutions of the above system of PDEs. Thus,
solving the above system of PDEs by standard methods found in the theory of
differential equations, we find that
\[
\left\{  f\left(  x\right)  =\sqrt{x_{1}^{2}+x_{2}^{2}+x_{3}^{2}}\right\}
\]
is a complete set of entanglement invariants. \ 

\bigskip

A functionally equivalent complete set of entanglement invariants is
\[
\left\{  f^{\prime}=x_{1}^{2}+x_{2}^{2}+x_{3}^{2}\right\}  \text{ ,}%
\]
which is also a basic set of entanglement invariants. \ 

\bigskip

\begin{remark}
Fortunately, in this simplest case, a complete set of entanglement invariants
and a basic set of entanglement invariants are one and the same. \ This will
not be the case for quantum systems of more than one qubit.
\end{remark}

\bigskip

\subsection{The Bloch ``sphere''}

\bigskip

As a result of the previous calculation, we have a complete set of
entanglement invariants, namely
\[
f\left(  x\right)  =\sqrt{x_{1}^{2}+x_{2}^{2}+x_{3}^{2}}=\left|  x\right|
\]
We have completely classified all the entanglement classes for 1 qubit quantum
systems. For in this case,
\[
\left[  i\rho\right]  _{E}=\left[  i\rho^{\prime}\right]  \Longleftrightarrow
f(i\rho)=f(i\rho^{\prime})\text{ .}%
\]

\bigskip

As a consequence of this result, the induced foliation of the space
$\mathbf{density}\left(  2^{1}\right)  $ of all physical density operators
lying in in the Lie algebra $\mathbf{u}\left(  2^{1}\right)  $ can be
visualized in terms of the 3-ball or radius 1 in $\mathbb{R}^{3}$, called the
\textbf{Bloch} ``\textbf{sphere}.'' \ 

\bigskip

Recall from remark on page \pageref{physical density operator} that
\[
\mathbf{density}\left(  2^{1}\right)  =\left\{  i\rho\in\mathbf{u}\left(
2^{1}\right)  \mid\rho\text{ is positive semi-definite and of trace
one}\right\}
\]
is a convex subset of the of the Lie algebra $\mathbf{u}\left(  2^{1}\right)
$. \ In this special case of $n=1$ qubit, it is a straight forward exercise to
show that
\[
\mathbf{density}\left(  2^{1}\right)  =\left\{  i\rho=x_{0}\xi_{0}+x\cdot
\xi\mid\left|  x\right|  \leq1\text{ and }x_{0}=-1\right\}  \text{ .}%
\]

\bigskip

Thus, the convex subset $\mathbf{density}\left(  2^{1}\right)  $ of
$\mathbf{u}\left(  2^{1}\right)  $ of all physical density operators $i\rho$
in $\mathbf{u}\left(  2^{1}\right)  $ can naturally be identified with the
3-ball of radius one via the one-to-one correspondence
\[
i\rho=x_{0}\xi_{0}+x_{1}\xi_{1}+x_{2}\xi_{2}+x_{3}\xi_{3}\longleftrightarrow
\left(  x_{1},x_{2},x_{3}\right)
\]
as illustrated in Figure 5.

\bigskip%

\begin{center}
\fbox{\includegraphics[
height=2.3722in,
width=2.9196in
]%
{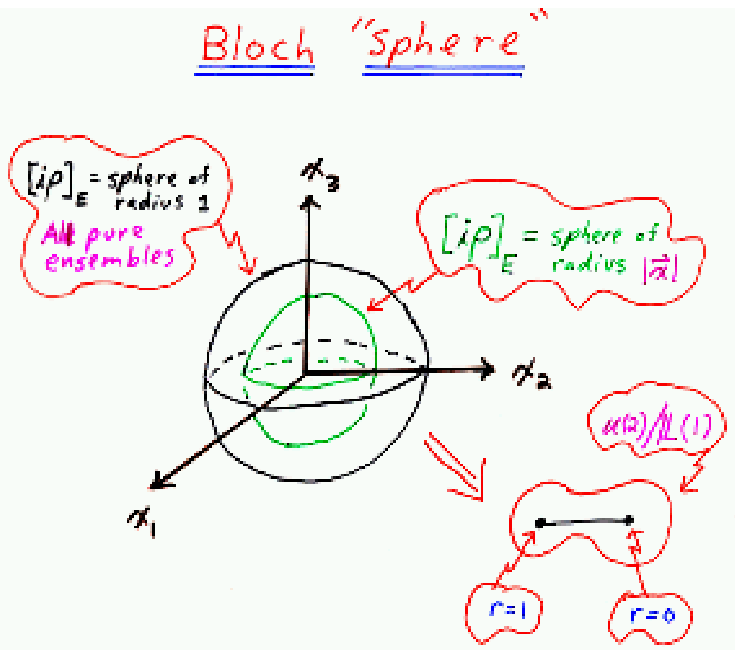}%
}\\
\textbf{Figure 5. \ The Bloch ``sphere''}%
\end{center}

\bigskip

It follows that each entanglement class $\left[  i\rho\right]  _{E}$ is simply
a sphere of radius $f(i\rho)=\left|  x\right|  $. \ The sphere of radius one
is the entanglement class of all pure ensembles. \ All other spheres represent
entanglement classes of mixed ensembles. \ The ``sphere'' of radius $0$ (i.e.,
the origin) represents the entanglement class of the maximally mixed ensemble. \ 

\bigskip

IAnd so we can conclude that the space of entanglement classes lying in the
space formed by identifying the elements of the convex set $\mathbf{density}%
\left(  2^{1}\right)  $ via the action of the local transformation group
$\mathbb{L}\left(  2^{1}\right)  $, namely
\[
\mathbf{density}\left(  2^{1}\right)  /\mathbb{L}\left(  2^{1}\right)  \text{
,}%
\]
is simply a closed\footnote{The adjective ``closed'' means that the line
segment contains both its endpoints.} line segment.

\bigskip

In terms of this picture, it is easy to visualize the tangent space $T_{i\rho
}\left(  \left[  i\rho\right]  _{E}\right)  $ to $\left[  i\rho\right]  _{E}$
at $i\rho$. \ Moreover, it is easy to visualize the normal bundle of $\left[
i\rho\right]  _{E}$. \ For the normal vector field is simply
\[
\left.  x_{1}\frac{\partial}{\partial x_{1}}+x_{2}\frac{\partial}{\partial
x_{2}}+x_{3}\frac{\partial}{\partial x_{3}}\right|  _{\left[  i\rho\right]
_{E}}%
\]

\bigskip

Unfortunately, for quantum systems of more than one qubit, such a
visualization is by no means as easy.

\bigskip

\section{Example 2. The entanglement classes of $n=2$ qubits}

\bigskip

As one might expect, the entanglement of two qubit quantum systems is much
more complex than that of one qubit quantum systems. \ In fact, with each
additional qubit, the entanglement becomes exponentially more complex than
before. \ Perhaps this is a strong hint as to where the power of quantum
computation is coming from?

\bigskip

For this example, the local unitary group $\mathbb{L}\left(  2^{2}\right)  $
is the Lie group $\mathbb{SU}\left(  2^{1}\right)  \otimes\mathbb{SU}\left(
2^{1}\right)  $. \ The corresponding Lie algebra $\mathbf{\ell}\left(
2^{2}\right)  $ is Kronecker sum\footnote{We remind the reader that the
Kronecker sum $A\boxplus B$ of two matrices (operators) $A$ and $B$ is defined
as
\[
A\boxplus B=A\otimes\mathbf{1}+\mathbf{1}\otimes B
\]
where $\mathbf{1}$ denotes the identity matrix (operator).} $\mathbf{su}%
(2)\boxplus\mathbf{su}(2)$. \ Each density operator $i\rho$ lies in the Lie
algebra $\mathbf{u}\left(  2^{2}\right)  $. \ 

\bigskip

As an immediate consequence of Proposition \ref{ad} given in Section
\ref{Section Act Def}, the infinitesimal action
\[
\Omega:\mathbf{\ell}\left(  2^{2}\right)  \longrightarrow\mathbf{Vec}\left(
\mathbf{u}\left(  2^{2}\right)  \right)
\]
is simply the small adjoint action, i.e.,
\[
\Omega\left(  v\right)  =ad_{v}\text{ ,}%
\]
for all $v\in\mathbf{\ell}\left(  2^{2}\right)  $.

\bigskip

We can now use the bases\footnote{See Section \ref{Section Lie Alg}.}
\[%
\begin{array}
[c]{c}%
\left\{  \xi_{10},\xi_{20},\xi_{30},\xi_{01},\xi_{02},\xi_{03},\right\} \\
\text{ \ and \ }\\
\left\{  \xi_{ij}\mid i,j=0,1,2,3\right\}
\end{array}
\]
of the respective Lie algebras $\mathbf{\ell}\left(  2^{2}\right)  $ and
$\mathbf{u}\left(  2^{2}\right)  $ to find a more useful expression for
$\Omega\left(  v\right)  $, where
\[
\xi_{ij}=-\frac{i}{2}\sigma_{i}\otimes\sigma_{j}\text{ .}%
\]
. \ 

\bigskip

Each element $v\in\mathbf{\ell}\left(  2^{2}\right)  $ can be uniquely
expressed in the form
\[
v=\left(  a\cdot\xi\right)  \otimes I_{4}+I_{4}\otimes\left(  b\cdot
\xi\right)  =a\cdot\xi\boxplus b\cdot\xi\text{ ,}%
\]
where $a=\left(  a_{1},a_{2},a_{3}\right)  $ and $b=\left(  b_{1},b_{2}%
,b_{3}\text{,}\right)  $ lie in $\mathbb{R}^{3}$, where $\xi=\left(  \xi
_{1},\xi_{2},\xi_{3}\right)  $, and where $I_{4}$ is the $4\times4$ identity
matrix. \ Thus,
\[%
\begin{array}
[c]{ccl}%
\Omega\left(  v\right)  & = & \Omega\left(  \sum_{j=1}^{3}\left(  a_{j}%
\xi_{j0}+b_{j}\xi_{0j}\right)  \right) \\
&  & \\
& = & \Omega\left(  a\cdot\xi\boxplus b\cdot\xi\right) \\
&  & \\
& = & ad_{a\cdot\xi\boxplus b\cdot\xi}\\
&  & \\
& = & ad_{\left(  a\cdot\xi\right)  \otimes I_{4}}+ad_{I_{4}\otimes\left(
b\cdot\xi\right)  }\\
&  & \\
& = & I_{4}\otimes\left(  a\cdot ad_{\xi}\right)  +\left(  b\cdot ad_{\xi
}\right)  \otimes I_{4}%
\end{array}
\]
where
\[
ad_{\xi}=\left(  ad_{\xi_{1}},ad_{\xi_{2}},ad_{\xi_{3}}\right)  \text{ .}%
\]

\bigskip

But as in example 1,
\[
ad_{\xi_{j}}=\left\{
\begin{array}
[c]{ll}%
\left(
\begin{array}
[c]{cc}%
0 & 0\\
0 & L_{j}%
\end{array}
\right)  =0\oplus L_{j} & \text{if }j=1,2,3\\
& \\
\left(
\begin{array}
[c]{cc}%
0 & 0\\
0 & 0
\end{array}
\right)  =0 & \text{if }j=0
\end{array}
\right.  \text{ ,}%
\]
where
\[
L_{1}=\left(
\begin{array}
[c]{rrr}%
0 & 0 & 0\\
0 & 0 & -1\\
0 & 1 & 0
\end{array}
\right)  \text{, }L_{2}=\left(
\begin{array}
[c]{rrr}%
0 & 0 & 1\\
0 & 0 & 0\\
-1 & 0 & 0
\end{array}
\right)  \text{, }L_{3}=\left(
\begin{array}
[c]{rrr}%
0 & -1 & 0\\
1 & 0 & 0\\
0 & 0 & 0
\end{array}
\right)
\]
is the basis of the Lie algebra $\mathbf{so}\left(  3\right)  $ of the special
orthogonal group $\mathbb{SO}\left(  3\right)  $ given in Appendix B on page
\pageref{so3 basis}.

\bigskip

\bigskip

Let
\[
\left\{  \partial/\partial x_{jk}\mid j,k=0,1,2,3\right\}
\]
denote the basis of $\mathbf{Vec}\left(  \mathbf{u}\left(  2^{2}\right)
\right)  $ induced by the chart
\[%
\begin{array}
[c]{ccc}%
\mathbf{u}\left(  2^{2}\right)  & \overset{\pi}{\longrightarrow} &
\mathbb{R}^{16}\\
&  & \\
i\rho=\sum_{i,j=0}^{3}x_{ij}\xi_{ij} & \longmapsto & \left(  x_{00},x_{0\ast
},x_{10},x_{1\ast},x_{20},x_{2\ast},x_{30},x_{3\ast}\right)
\end{array}
\]
where
\[%
\begin{array}
[c]{l}%
\left(  x_{00},x_{0\ast},x_{10},x_{1\ast},x_{20},x_{2\ast},x_{30},x_{3\ast
}\right) \\
\\
=\left(  x_{00},\quad x_{01},x_{02},x_{03},\quad x_{10},\quad x_{11}%
,x_{12},x_{13},\quad x_{20},\quad x_{21},x_{22},x_{23},\quad x_{30},\quad
x_{31},x_{32},x_{33}\right)
\end{array}
\]

\bigskip

In other words, for each pair $\left(  j,k\right)  $, $\partial/\partial
x_{jk}$ denotes the vector field on $\mathbf{u}\left(  2^{2}\right)  $ defined
at each point $i\rho$ as the tangent vector to the curve
\[
\pi^{-1}\left(  x_{00},\ldots,x_{jk}+t,\ldots,x_{33}\right)  =i\rho+t\xi_{jk}%
\]
at $t=0$.

\bigskip

In terms of the above chart, $\Omega\left(  v\right)  \left(  i\rho\right)  $
can be written as
\[
\left(  x_{00},x_{0\ast},x_{10},x_{1\ast},x_{20},x_{2\ast},x_{30},x_{3\ast
}\right)  \cdot\left[  I_{4}\otimes\left(  0\oplus a\cdot L\right)  +\left(
0\oplus b\cdot L\right)  \otimes I_{4}\right]  \cdot\left(
\begin{array}
[c]{c}%
\partial/\partial x_{00}\\
\partial/\partial x_{0\ast}\\
\partial/\partial x_{10}\\
\partial/\partial x_{1\ast}\\
\partial/\partial x_{20}\\
\partial/\partial x_{2\ast}\\
\partial/\partial x_{30}\\
\partial/\partial x_{3\ast}%
\end{array}
\right)  \text{ ,}%
\]
which simplifies to
\[
\Omega\left(  v\right)  \left(  i\rho\right)  =\sum_{q=0}^{3}\left(  a\cdot
x_{q\ast}\times\frac{\partial}{\partial x_{q\ast}}+b\cdot x_{\ast q}%
\times\frac{\partial}{\partial x_{\ast q}}\right)  \text{ ,}%
\]
where `$\times$' denotes the vector cross product\footnote{The vector cross
product is computed according to the right-hand rule.}.

\bigskip

We can now achieve objective 1.

\bigskip

\noindent\textbf{Objective 1.} \ \textit{Given an arbitrary density operator}
$i\rho$ \textit{in} $\mathbf{u}\left(  2\right)  $, \textit{find the dimension
of an arbitrary entanglement class} $\left[  i\rho\right]  _{E}$.

\bigskip

From the above discussion, it follows that the image $\operatorname{Im}\left(
\Omega\right)  $ of the infinitesimal action $\Omega$ is spanned by the six
vector fields

\bigskip%

\[%
\begin{array}
[c]{c}%
\left\{
\begin{array}
[c]{c}%
\Omega\left(  \xi_{01}\right)  =\sum\limits_{q=0}^{3}\left(  x_{q2}%
\frac{\partial}{\partial x_{q3}}-x_{q3}\frac{\partial}{\partial x_{q2}}\right)
\\
\\
\Omega\left(  \xi_{02}\right)  =\sum\limits_{q=0}^{3}\left(  x_{q3}%
\frac{\partial}{\partial x_{q1}}-x_{q1}\frac{\partial}{\partial x_{q3}}\right)
\\
\\
\Omega\left(  \xi_{03}\right)  =\sum\limits_{q=0}^{3}\left(  x_{q1}%
\frac{\partial}{\partial x_{q2}}-x_{q2}\frac{\partial}{\partial x_{q1}%
}\right)
\end{array}
\right. \\
\\
\left\{
\begin{array}
[c]{c}%
\Omega\left(  \xi_{10}\right)  =\sum\limits_{q=0}^{3}\left(  x_{2q}%
\frac{\partial}{\partial x_{3q}}-x_{3q}\frac{\partial}{\partial x_{2q}}\right)
\\
\\
\Omega\left(  \xi_{20}\right)  =\sum\limits_{q=0}^{3}\left(  x_{3q}%
\frac{\partial}{\partial x_{1q}}-x_{1q}\frac{\partial}{\partial x_{3q}}\right)
\\
\\
\Omega\left(  \xi_{30}\right)  =\sum\limits_{q=0}^{3}\left(  x_{1q}%
\frac{\partial}{\partial x_{2q}}-x_{2q}\frac{\partial}{\partial x_{1q}%
}\right)
\end{array}
\right.
\end{array}
\]

In particular, the tangent space $T_{i\rho}\left(  \left[  i\rho\right]
_{E}\right)  =\left.  \left(  \operatorname{Im}\Omega\right)  \right|
_{i\rho}$ to the entanglement class $\left[  i\rho\right]  _{E}$ at the point
$i\rho$ is spanned by
\[
\left.  \Omega\left(  \xi_{01}\right)  \right|  _{i\rho}\text{, }\left.
\Omega\left(  \xi_{02}\right)  \right|  _{i\rho}\text{, }\left.  \Omega\left(
\xi_{03}\right)  \right|  _{i\rho}\text{, }\left.  \Omega\left(  \xi
_{10}\right)  \right|  _{i\rho}\text{, }\left.  \Omega\left(  \xi_{20}\right)
\right|  _{i\rho}\text{, }\left.  \Omega\left(  \xi_{30}\right)  \right|
_{i\rho}%
\]
We leave it as an exercise for the reader to verify that the above six vector
fields are linearly independent almost every where. Thus, it follows that
almost all entanglement classes are of dimension six.

\bigskip

However, there are notable exceptions. \ Consider the Bell state\footnote{It
should be noted that all four 2 qubit Bell states lie in the same entanglement
class. \ It is this fact that makes quantum teleportation possible.}, $\left|
\psi\right\rangle =\frac{1}{\sqrt{2}}\left(  \left|  00\right\rangle -\left|
11\right\rangle \right)  $. \ The corresponding density operator $i\rho$ is
\[
i\rho=\frac{i}{2}\left(
\begin{array}
[c]{rrrr}%
1 & 0 & 0 & -1\\
0 & 0 & 0 & 0\\
0 & 0 & 0 & 0\\
-1 & 0 & 0 & 1
\end{array}
\right)  =\left(  -\frac{1}{2}\right)  \xi_{00}+\left(  \frac{1}{2}\right)
\xi_{11}+\left(  -\frac{1}{2}\right)  \xi_{22}+\left(  -\frac{1}{2}\right)
\xi_{33}\text{ ,}%
\]
where
\[
\xi_{jk}=-\frac{i}{2}\sigma_{j}\otimes\sigma_{k}\text{ .}%
\]
Hence,
\[
x_{jk}=\left\{
\begin{array}
[c]{rcl}%
-\frac{1}{2} & \text{if} & j=k=0,2,3\\
&  & \\
\frac{1}{2} & \text{if} & j=k=1\\
&  & \\
0 & \text{if} & j\neq k
\end{array}
\right.
\]

\bigskip

Thus, in this case $\left.  \operatorname{Im}\left(  \Omega\right)  \right|
_{i\rho}$ is spanned by%

\[%
\begin{array}
[c]{c}%
\left\{
\begin{array}
[c]{lll}%
\left.  \Omega\left(  \xi_{01}\right)  \right|  _{i\rho} & = & \frac{1}%
{2}\left(  \frac{\partial}{\partial x_{23}}-\frac{\partial}{\partial x_{32}%
}\right) \\
&  & \\
\left.  \Omega\left(  \xi_{02}\right)  \right|  _{i\rho} & = & \frac{1}%
{2}\left(  \frac{\partial}{\partial x_{31}}+\frac{\partial}{\partial x_{13}%
}\right) \\
&  & \\
\left.  \Omega\left(  \xi_{03}\right)  \right|  _{i\rho} & = & \frac{1}%
{2}\left(  -\frac{\partial}{\partial x_{12}}-\frac{\partial}{\partial x_{21}%
}\right)
\end{array}
\right. \\
\\
\left\{
\begin{array}
[c]{lll}%
\left.  \Omega\left(  \xi_{10}\right)  \right|  _{i\rho} & = & \frac{1}%
{2}\left(  \frac{\partial}{\partial x_{32}}-\frac{\partial}{\partial x_{23}%
}\right) \\
&  & \\
\left.  \Omega\left(  \xi_{20}\right)  \right|  _{i\rho} & = & \frac{1}%
{2}\left(  \frac{\partial}{\partial x_{13}}+\frac{\partial}{\partial x_{31}%
}\right) \\
&  & \\
\left.  \Omega\left(  \xi_{30}\right)  \right|  _{i\rho} & = & \frac{1}%
{2}\left(  -\frac{\partial}{\partial x_{21}}-\frac{\partial}{\partial x_{12}%
}\right)
\end{array}
\right.
\end{array}
\]

\bigskip

Hence,
\[
Dim\left[  i\rho_{Bell}\right]  _{E}=Dim\left[  \left.  \left(
\operatorname{Im}\Omega\right)  \right|  _{i\rho_{Bell}}\right]  =3
\]
This only confirms the conventional wisdom that the entanglement class of the
Bell states is truly exceptional.

\bigskip

We are now ready for objective 2:

\bigskip

\noindent\textbf{Objective 2.} \ \textit{Given two states} $i\rho$
\textit{and} $i\rho^{\prime}$, \textit{devise a means of determining whether
they belong to the same or to different entanglement classes}. \ 

\bigskip

The complete functionally independent set of solutions to the above system of
PDEs (hence, a complete set of entanglement invariants) was found by Linden
and Popescu in\cite{Linden2}. \ These invariants are as described
below\footnote{We are using a notation different from that found in
\cite{Linden2}.}. \ For further details please refer to \cite{Linden2}. \ 

\bigskip

Let $i\rho$ be an arbitrary element of the Lie algebra $\mathbf{u}\left(
2^{2}\right)  $. \ Then in terms of the earlier described chart $\pi$,
\[
i\rho=\sum_{j,k=0}^{3}x_{jk}\xi_{jk}\text{ ,}%
\]
where $\left\{  \xi_{jk}\right\}  $ denotes the basis of $\mathbf{u}\left(
2^{2}\right)  $ described earlier.

\bigskip

We will change our notation slightly. \ Let $x_{\ast\ast}$ denote the
$3\times3$ matrix
\[
x_{\ast\ast}=\left(  x_{jk}\right)  _{j,k=1,2,3}\text{ ,}%
\]
and let $x_{0\ast}$ and $x_{\ast0}$ denote the vectors
\[
\left\{
\begin{array}
[c]{ccc}%
x_{0\ast} & = & \left(  x_{01},x_{02},x_{02}\right) \\
&  & \\
x_{\ast0} & = & \left(  x_{10},x_{20},x_{30}\right)
\end{array}
\right.
\]
Finally, let $Z$ denote the matrix
\[
Z=x_{\ast\ast}x_{\ast\ast}^{T}\text{ ,}%
\]
where the superscript `$T$' denotes the transpose. \ Then the nine
algebraically independent polynomial functions listed in the table
\[%
\begin{tabular}
[c]{||c||c||c||}\hline\hline
$Tr\left(  Z\right)  $ & $\overset{}{\underset{}{Tr\left(  Z^{2}\right)  }}$ &
$\det\left(  x_{\ast\ast}\right)  $\\\hline\hline
$x_{0\ast}x_{0\ast}^{T}$ & $\overset{}{\underset{}{x_{0\ast}Zx_{0\ast}^{T}}}$
& $x_{0\ast}Z^{2}x_{0\ast}^{T}$\\\hline\hline
$x_{0\ast}x_{\ast\ast}x_{\ast0}^{T}$ & $\overset{}{\underset{}{x_{0\ast
}Zx_{\ast\ast}x_{\ast0}^{T}}}$ & $x_{0\ast}Z^{2}x_{\ast\ast}x_{\ast0}^{T}%
$\\\hline\hline
\end{tabular}
\]
form a basic set of entanglement invariants.

\bigskip

\textit{But the above nine entanglement invariants do not form a complete set
of entanglement invariants}! \ A tenth polynomial function
\[
x_{0\ast}\cdot\left(  Zx_{0\ast}^{T}\right)  \times\left(  Z^{2}x_{0\ast}%
^{T}\right)  \text{ }%
\]
is needed to form a complete system of entanglement invariants. \ Although
this tenth entanglement invariant is algebraically dependent on the above nine
entanglement invariants, it is still needed to determine the sign of the
components of $i\rho$.

\bigskip

\section{Example $n$. The entanglement classes of $n$ qubits, $n>2$}

\qquad\bigskip

For $n$ qubits ($n>2$), the same methods lead to the following formula for the
infinitesimal action
\[
\fbox{$%
\begin{array}
[c]{c}%
\Omega\left(  v\right)  \left(  i\rho\right)  =%
{\displaystyle\sum\limits_{q_{1},q_{2}\cdots q_{n-1}=0}^{3}}
{\displaystyle\sum\limits_{k=1}^{n}}
a^{(k)}\cdot x_{q_{1}q_{2}\cdots q_{k-1}\ast q_{k+1}\cdots q_{n-1}}%
\times\frac{\partial}{\partial x_{q_{1}q_{2}\cdots q_{k-1}\ast q_{k+1}\cdots
q_{n-1}}}%
\end{array}
$}%
\]

\bigskip

\noindent where $v\in\mathbf{\ell}\left(  2^{n}\right)  $ and $i\rho
\in\mathbf{u}\left(  2^{n}\right)  $ are given by
\[
\left\{
\begin{array}
[c]{ccl}%
v & = &
{\displaystyle\sum\limits_{k=1}^{n}}
a^{(k)}\cdot\xi\underset{\ast\text{ in }k\text{-th position}}{_{\underbrace
{00\cdots0\ast0\cdots0}}}\\
&  & \\
i\rho & = &
{\displaystyle\sum\limits_{r_{1},r_{2},\cdots,r_{n}=0}^{3}}
x_{r_{1}r_{2}\cdots r_{n}}\xi_{r_{1}r_{2}\cdots r_{n}}%
\end{array}
\right.
\]

\bigskip

We will leave the solution to the corresponding system of PDEs to future papers.

\bigskip

\section{Conclusion}

\bigskip

There is much more that could be said about quantum entanglement. \ This paper
presents only a small part of the big picture. \ But hopefully this paper will
provide the reader with some insight into this rapidly growing research field.
\ Since this paper was written, research in quantum entanglement has literally
had an explosive expansion, and even now continues to do so. \ We refer the
reader to the references at the end of this paper, which represent only a few
of the many papers in this rapidly expanding field.

\bigskip

\section{Appendix A. \ Some Fundamental Concepts from the Theory of
Differential Manifolds}

\bigskip

\subsection{Differential manifolds, tangent bundles, and vector fields}

\bigskip

\begin{definition}
A topological space $M^{m}$ is an $\mathbf{m}$\textbf{-dimensional manifold}
if it is locally homeomorphic to $\mathbb{R}^{m}$, i.e., if there exists an
open cover $\mathcal{W}=\left\{  W_{\alpha}\right\}  $ of $M^{m}$ such that
for each $W_{\alpha}\in\mathcal{W}$, there is associated a homeomorphism
\begin{gather*}
W_{\alpha}\overset{\varphi_{\alpha}}{\longrightarrow}\mathbb{R}^{m}\\
x\longmapsto\left(  x_{1},x_{2},\ldots,x_{m}\right)
\end{gather*}
which maps $W_{\alpha}$ onto an open subset of $\mathbb{R}^{m}$. \ We call
\[
\left(  \varphi_{\alpha},W_{\alpha}\right)
\]
a \textbf{chart on }$M^{m}$, and
\[
\Phi=\left\{  \ \left(  \varphi_{\alpha},W_{\alpha}\right)  \ \right\}
\]
an \textbf{Atlas} on $M^{m}$.

An Atlas is said to be \textbf{smooth} ($C^{\infty}$), if whenever
\[
\varphi_{\beta}\varphi_{\alpha}^{-1}:\varphi_{\alpha}\left(  W_{\alpha}\cap
W_{\beta}\right)  \longrightarrow\varphi_{\beta}\left(  W_{\alpha}\cap
W_{\beta}\right)
\]
is defined, is a smooth ($C^{\infty}$) map \ of $\varphi_{\alpha}\left(
W_{\alpha}\cap W_{\beta}\right)  \subseteq\mathbb{R}^{n}$ into $\varphi
_{\beta}\left(  W_{\alpha}\cap W_{\beta}\right)  \subseteq\mathbb{R}^{n}$. \ A
\textbf{smooth }($C^{\infty}$) \textbf{manifold }is a topological manifold
with a smooth atlas.
\end{definition}

\bigskip

\textbf{%
\begin{center}
\fbox{\includegraphics[
height=1.2263in,
width=3.6971in
]%
{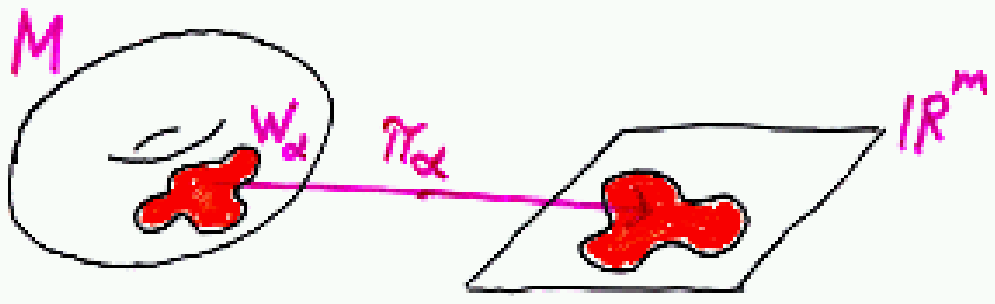}%
}\\
\textbf{Figure 6. \ A chart }$\mathbf{\pi}_{\alpha}\mathbf{:M\longrightarrow
}\mathbb{R}^{\mathbf{4}}$\textbf{\ on a manifold }$\mathbf{M}$\textbf{.}%
\end{center}
}

\bigskip%

\begin{center}
\fbox{\includegraphics[
height=2.2572in,
width=2.636in
]%
{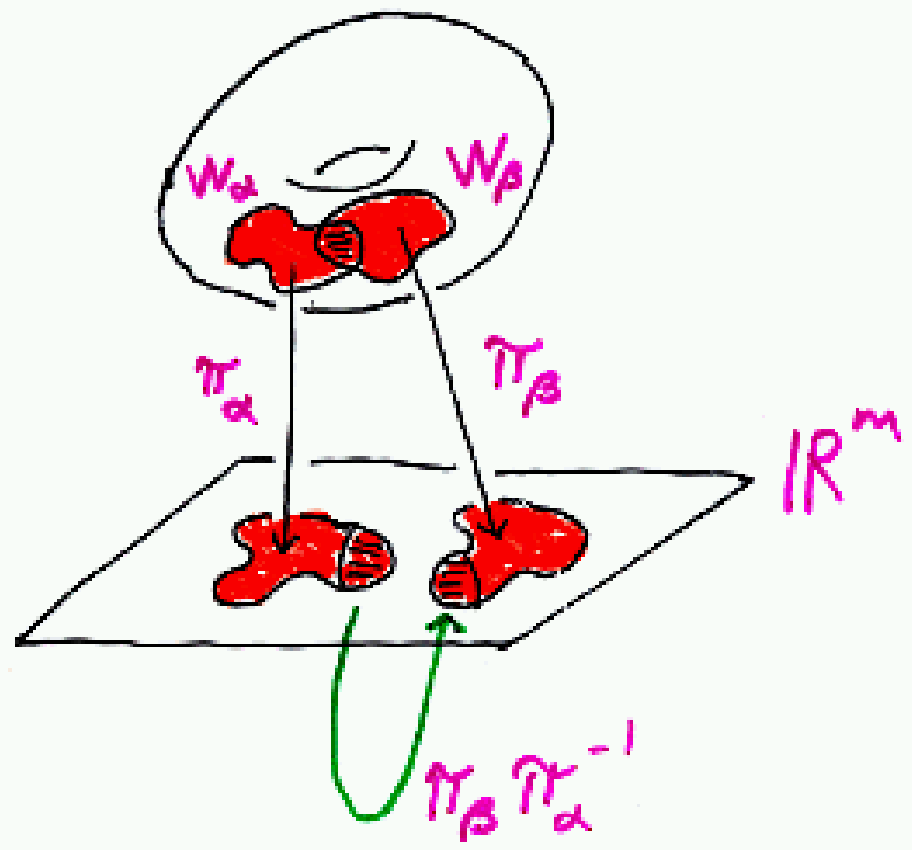}%
}\\
\textbf{Figure 7. \ An atlas is smooth if every} $\mathbf{\pi}_{\alpha
}\mathbf{\pi}_{\beta}^{-1}$ \textbf{is smooth when defined.}%
\end{center}

\bigskip

\bigskip

\begin{definition}
Let $M$ and $N$ be smooth manifolds. \ Then a map
\[
f:M\longrightarrow N
\]
is said to be smooth if for every $x\in M$ there exist charts $\left(
\varphi_{\alpha},W_{\alpha}\right)  $ of $M$ and $\left(  \psi_{\beta
},V_{\beta}\right)  $ of $N$ containing $x$ and $f(x)$ respectively such that
\[
\psi_{\beta}f\varphi_{\alpha}^{-1}:\varphi_{\alpha}\left(  W_{\alpha}\right)
\longrightarrow\varphi_{\beta}\left(  V_{\beta}\right)
\]
is smooth.
\end{definition}

\bigskip

\begin{definition}
Let x be an element of a smooth manifold $M$, and let $\gamma_{1}(t)$ and
$\gamma_{2}(t)$ be smooth curves in $M$ which pass through $x$, i.e., such
that there exists $t_{1},t_{2}\in\mathbb{R}$ for which
\[
\gamma_{1}(t_{1})=x=\gamma_{2}(t_{2})
\]
Then $\gamma_{1}$ and $\gamma_{2}$ are said to be \textbf{tangentially
equivalent} at $x$, written
\[
\gamma_{1}\underset{x}{\thicksim}\gamma_{2}\text{ ,}%
\]
if they are tangent at the point $x$, i.e., if there is a chart $\left(
\varphi_{\alpha},W_{\alpha}\right)  $ on $M$ containing $x$ such that
\[
\frac{d}{dt}\left(  \varphi_{\alpha}\circ\gamma_{1}\right)  (t)\mid_{t=t_{1}%
}=\frac{d}{dt}\left(  \varphi_{\alpha}\circ\gamma_{2}\right)  (t)\mid
_{t=t_{2}}%
\]
\end{definition}

\bigskip

\begin{remark}
It can easily be shown that the relation $\underset{x}{\thicksim}$\ is
independent of the chart selected.
\end{remark}

\bigskip

\begin{definition}
A\textbf{\ tangent vector} $\left(  x,v\right)  $ \ (also written simply as
$v$) to $M$ at $x$ is a tangential equivalence class at $x$. \ The tangent
space of $M^{n}$ at $x$, denoted by $T_{x}M^{n}$, is the set of tangent
vectors to $M$ at $x$. \ $T_{x}M$ can be shown to be an $n$-dimensional vector space.

Let
\[
TM=%
{\displaystyle\bigcup\limits_{x\in M}}
T_{x}M\text{ ,}%
\]
and let $\pi$ be the map
\begin{gather*}
TM\overset{\pi}{\longrightarrow}M\\
(x,v)\longmapsto x
\end{gather*}
If $\varphi_{\alpha}:W_{\alpha}\longrightarrow\mathbb{R}^{m}$ is a chart on
$M$, then
\[
\varphi_{\alpha}\pi:\pi^{-1}W_{\alpha}\longrightarrow\mathbb{R}^{m}%
\]
can be shown to be a chart on $TM$. In this way, $TM$ becomes a smooth
manifold and $\pi$ becomes a smooth map. \ $TM$ together with the map $\pi$ is
called the \textbf{tangent bundle} of $M$. \ 
\end{definition}

\bigskip

\begin{definition}
A \textbf{vector field} $v$ on a smooth manifold $M^{m}$ is a smooth map
\[
v:M^{m}\longrightarrow TM^{m}%
\]
Let $\mathbf{Vec}(M^{m})$ be the set of all vector fields on the smooth
manifold $M^{m}$. \ This is easily seen to be a vector space where, for
example, the sum $u+v$ of two vector fields is defined by
\[
\left.  (u+v)\right|  _{x}=\left.  u\right|  _{x}+\left.  v\right|  _{x}\text{
}%
\]
for all $x\in M^{m}$.
\end{definition}

\bigskip

We will now consider the charts of the tangent bundle $TM$ in a more explicit way.

\bigskip

Let
\begin{gather*}
W_{\alpha}\overset{\varphi_{\alpha}}{\longrightarrow}\mathbb{R}^{m}\\
x\longmapsto\left(  x_{1},x_{2},\ldots,x_{m}\right)
\end{gather*}
be a chart on the smooth manifold $M^{m}$, and let $a$ be an arbitrary point
in $U_{\alpha}$. \ Thus,
\[
\varphi_{\alpha}(a)=\left(  a_{1},a_{2},\ldots,a_{m}\right)  \text{ .}%
\]

\bigskip

For each $j$ ($j=1,2,\ldots,m$) consider the smooth curve
\[
\gamma_{j}(t)=\varphi_{\alpha}^{-1}\left(  a_{1},a_{2},\ldots,a_{j}%
+t,\ldots,a_{m}\right)
\]
in $U_{\alpha}$ which passes through the point $a$ at time $t=0$. Then for
each such $j$, let \label{vec basis}
\[
\left.  \frac{\partial}{\partial x_{j}}\right|  _{a}\in T_{a}M
\]
denote the tangent vector to the curve $\gamma_{j}$ at $a$. \ It can be shown
that
\[
\left.  \frac{\partial}{\partial x_{1}}\right|  _{a}\text{, }\left.
\frac{\partial}{\partial x_{2}}\right|  _{a}\text{, }\ldots\text{, }\left.
\frac{\partial}{\partial x_{m}}\right|  _{a}%
\]
is a vector space basis of the tangent space $T_{a}M$. \ 

\bigskip

Moreover, since this construction is respect to an arbitrary point $a$ in
$W_{\alpha}$, it can be shown that we have actually constructed for each $j$ a
smooth vector field
\[
\frac{\partial}{\partial x_{j}}\in\mathbf{Vec}(TW_{\alpha})\subseteq
\mathbf{Vec}(TM)
\]
In fact, it can be shown that
\[
\frac{\partial}{\partial x_{1}}\text{, }\frac{\partial}{\partial x_{2}}\text{,
}\ldots\text{, }\frac{\partial}{\partial x_{m}}%
\]
is a basis of $\mathbf{Vec}(TW_{\alpha})$, and hence a local basis of
$\mathbf{Vec}(TM)$.

\bigskip

We can now express each chart $\left(  \varphi_{\alpha}\pi,\pi^{-1}W_{\alpha
}\right)  $ explicitly as:
\begin{gather*}
\qquad\pi^{-1}W_{\alpha}\overset{\varphi_{\alpha}\pi}{\longrightarrow
}\mathbb{R}^{2m}\\
(x,\mu_{1}\frac{\partial}{\partial x_{1}}+\mu_{2}\frac{\partial}{\partial
x_{2}}+\ldots+\mu_{m}\frac{\partial}{\partial x_{m}})\longmapsto\left(
x_{1},x_{2},\ldots,x_{m},\ \mu_{1},\mu_{2},\ldots,\mu_{m}\right)
\end{gather*}
where $\mu_{j}$'s on the left denote functions of $x\in M$, and where $\mu
_{j}$'s on the right denote functions of $\left(  x_{1},x_{2},\ldots
,x_{m}\right)  \in\mathbb{R}^{m}$. \bigskip

\begin{definition}
Let $M$ and $N$ be smooth manifolds, and let $f:M\longrightarrow N$ be a
smooth map, and let$\quad a$\quad be an arbitrary point of $M$. \ We define a
vector space morphism
\[
\left.  df\right|  _{a}:T_{a}M\longrightarrow T_{f(a)}N
\]
as follows:

For each $v\in T_{a}M$, there is a representative smooth curve $\gamma_{v}(t)$
in $M$ which passes through the point $a$ and which has $v$ as its tangent
vector at the point $a$. \ It follows that $f\circ\gamma_{v}(t)$ is a smooth
curve in $N$ passing through the point $f(a)$. \ We define
\[
\left.  df\right|  _{a}(v)\in T_{f(a)}N
\]
as the tangent vector to $f\circ\gamma_{v}(t)$ at the point $f(a)$. \ It is
then a simple exercise to show that $\left.  df\right|  _{a}$ is a vector
space morphism.

Since $a$ was an arbitrary point of $M$, this leads to the definition of a
smooth map $df:TM\longrightarrow TN$, called the differential of $f$, such
that the following diagram is commutative:
\[%
\begin{array}
[c]{cccc}%
TM & \overset{df}{\longrightarrow} & TN & \\
\downarrow &  & \downarrow & \\
M & \overset{f}{\longrightarrow} & N & \text{.}%
\end{array}
\]
\end{definition}

\bigskip

\begin{remark}
In local coordinates, $df$ maps the tangent vector
\[
\left.  v\right|  _{x}=\sum_{i=1}^{m}\mu_{i}\frac{\partial}{\partial x_{i}}%
\]
to the tangent vector
\[
df(\left.  v\right|  _{x})=\sum_{j=1}^{n}\left(  \sum_{i=1}^{m}\mu
_{i}\frac{\partial f_{j}}{\partial x_{i}}\right)  \frac{\partial}{\partial
y_{j}}\text{ .}%
\]
Thus, the matrix expression of the linear transformation $df$ is just the
Jacobian matrix
\[
\left(  \frac{\partial f_{j}}{\partial x_{i}}(x)\right)  _{m\times n}\text{
\ .}%
\]
\end{remark}

\bigskip

\subsection{Exponentiation of vector fields}

\quad\bigskip

\begin{definition}
Let $M$ be a smooth manifold, and let $v\in\mathbf{Vec}(M)$ be a smooth vector
field on $M$. \ A curve $\gamma(t)$ in $M$ is said to be an \textbf{integral
curve} of $v$ if $\left.  v\right|  _{\gamma(t)}$ is the tangent vector to
$\gamma(t)$ for each $t$ for which $\gamma(t)$ is defined. \ 
\end{definition}

In terms of local coordinates, an integral curve $\gamma(t)$ of a smooth
vector field
\[
v(x)=\sum_{i=1}^{n}\mu_{i}(x_{1},x_{2},\ldots,x_{m})\frac{\partial}{\partial
x_{i}}%
\]
is a solution to the system of ordinary differential equations
\[
\frac{dx_{i}}{dt}=\mu_{i}(x_{1},x_{2},\ldots,x_{m})\text{ , }i=1,2,\ldots,m
\]
Since $v$ is smooth, its coefficients $\mu^{i}(x_{1},x_{2},\ldots,x_{m})$ are
smooth functions. \ Consequently, it follows from the standard existence and
uniqueness theorems for systems of ordinary differential equations that there
exists a unique solution for each set of initial conditions.

\bigskip

Thus, for each $x$ in $M$, there exists a unique maximal integral curve
$\gamma_{v}(t,x)$ passing through $x$ at time $t=0$, and call $\gamma
_{v}(t,x)$ the \textbf{flow} generated by the vector field $v$. \ We call $v$
the \textbf{infinitesimal generator} of the flow. \ It can be easily shown
that
\[
\gamma_{v}(t,\gamma_{v}(s,x))=\gamma_{v}(t+s,x)\text{ \ .}%
\]
Hence, we are justified in adopting the following suggestive notation:
\[
e^{tv}x
\]
for the flow $\gamma_{v}(t,x)$.

\bigskip

In terms of our new notation, the properties of the flow can be expressed as

\begin{itemize}
\item[1)] $e^{sv}e^{tv}x=e^{(s+t)v}x$

\item[2)] $e^{0\cdot v}x=x$

\item[3)] $\frac{d}{dt}\left(  e^{tv}x\right)  =\left.  v\right|  _{e^{tv}x}$ \ .
\end{itemize}

\bigskip

\subsection{Vector fields viewed as directional derivatives}

\bigskip

We now show how vector fields can be viewed as partial differential operators.\bigskip

\begin{definition}
A \textbf{derivation} $D$ on an algebra $\mathcal{A}$ is a map
\[
D:\mathcal{A}\rightarrow\mathcal{A}%
\]
such that

\begin{itemize}
\item[1)] (Linearity) $D\left(  \alpha f+\beta g\right)  =\alpha Df+\beta Dg$

\item[2)] (Leibnitz Rule) $D\left(  fg\right)  =\left(  Df\right)  g+f\left(
Dg\right)  $
\end{itemize}
\end{definition}

\bigskip

\begin{definition}
A \textbf{Lie algebra} $\mathbb{A}$ is a vector space together with a binary
operation
\[
\left[  -,-\right]  :\mathbb{A}\times\mathbb{A\rightarrow A}\text{ ,}%
\]
called a \textbf{Lie bracket} for $\mathbb{A}$, such that

\begin{itemize}
\item[1)] (Bilinearity)
\[%
\begin{array}
[c]{ccc}%
\left[  \lambda_{1}a_{1}+\lambda_{2}a_{2},\ b\right]  & = & \lambda_{1}
\left[  a_{1},b\right]  +\lambda_{2}\left[  a_{2},b\right] \\
&  & \\
\left[  a,\ \lambda_{1}b_{1}+\lambda_{2}b_{2}\right]  & = & \lambda_{1}
\left[  a,b_{1}\right]  +\lambda_{2}\left[  a,b_{2}\right]
\end{array}
\]

\item[2)] (Skew-Symmetry)
\[
\left[  a,b\right]  =-\left[  b,a\right]
\]

\item[3)] (Jacobi Identity)
\[
\left[  a,\left[  b,c\right]  \right]  +\left[  c,\left[  a,b\right]  \right]
+\left[  b,\left[  c,a\right]  \right]  =0
\]
\end{itemize}
\end{definition}

\bigskip

\begin{proposition}
The set of derivations $Der(\mathcal{A})$ on an algebra $\mathcal{A}$ is a Lie
algebra with Lie bracket given by:
\[
\left[  D_{1},D_{2}\right]  =D_{1}\circ D_{2}-D_{2}\circ D_{1}%
\]
\end{proposition}

\bigskip

Let $\mathbf{C}^{\infty}(M)$ denote the \textbf{algebra of real valued
functions} on the smooth manifold $M$. \ Then, it follows that $\mathbf{Der}%
(\mathbf{C}^{\infty}(M))$ is a Lie algebra.

\bigskip

We will now show how to identify the elements of $\mathbf{Vec}(M)$ with
derivations in $\mathbf{Der}(\mathbf{C}^{\infty}(M))$, and thereby show that
$\mathbf{Vec}(M)$ is more than a vector space. \ It is actually a Lie algebra.\ 

\bigskip

Each smooth vector field $v$ on $M$ can be thought of as a directional
derivative in the direction $v$ as follows: \ Let $v\in\mathbf{Vec}(M)$ and
let $f\in\mathbf{C}^{\infty}(M)$. \ Define $v(f)$ as:
\[
\left.  v(f)\right|  _{x}=\left.  \frac{d}{dt}f\left(  e^{tv}x\right)
\right|  _{t=0}%
\]
\bigskip

Thus, we have:

\bigskip

\begin{proposition}
$\mathbf{Vec}(M)$ is a Lie algebra of derivations on the algebra
$\mathbf{C}^{\infty}(M)$.
\end{proposition}

\bigskip

It is enlightening, to view the above in terms of local coordinates. \ From
this perspective,
\[
v=\sum_{i=1}^{m}\mu_{i}(x)\frac{\partial}{\partial x_{i}}\text{ .}%
\]
Thus, if we use the chain rule and the fact that
\[
\frac{d}{dt}\left(  e^{tv}x\right)  =\left.  v\right|  _{e^{tv}x}\text{ ,}%
\]
we have
\[
\frac{d}{dt}f\left(  e^{tv}x\right)  =\sum_{i-1}^{m}\xi^{i}\left(
e^{tv}x\right)  \frac{\partial f}{\partial x^{i}}\left(  e^{tv}x\right)
=\left.  \left(  \sum_{i=1}^{m}\xi^{i}\frac{\partial}{\partial x^{i}}\right)
\left(  f\right)  \right|  _{e^{tv}x}\text{ .}%
\]
Hence,
\[
v=\sum_{i=1}^{m}\mu_{i}\frac{\partial}{\partial x_{i}}\in Vec(U_{\alpha
})\subseteq Vec(M)
\]
acts as a first order partial differential operator, thereby justifying the notation.

\bigskip

So viewing $v$ as a first order partial differential operator, we can write
\[
\frac{d}{dt}f\left(  e^{tv}x\right)  =\left.  v(f)\right|  _{e^{tv}x}\text{ ,}%
\]
and, in particular,
\[
\left.  \frac{d}{dt}f\left(  e^{tv}x\right)  \right|  _{t=0}=v(f)(x)\text{ ,}%
\]
where $v(f)(x)$ now denotes (locally) $\left(  \sum_{i=1}^{m}\mu
_{i}\frac{\partial}{\partial x_{i}}\right)  f$ evaluated at $x$.

\bigskip

\section{Appendix B. \ Some Fundamental Concepts from the Theory of Lie Groups}

\bigskip

\subsection{Lie groups}

\bigskip

\begin{definition}
A \textbf{Lie group} $\mathbb{G}$ is a group which is a smooth manifold whose
differential structure is compatible with the group operations, i.e., such that

\begin{itemize}
\item[1)] The multiplication map of $\mathbb{G}$
\begin{gather*}
\mathbb{G}\times\mathbb{G}\longrightarrow\mathbb{G}\\
\left(  g_{1},g_{2}\right)  \longmapsto g_{1}g_{2}%
\end{gather*}
and,

\item[2)] The inverse map of $G$
\begin{gather*}
\mathbb{G}\longrightarrow\mathbb{G}\\
g\longmapsto g^{-1}%
\end{gather*}
are smooth functions.
\end{itemize}

A closed subgroup $\mathbb{H}$ of $\mathbb{G}$ can be shown to be a subgroup,
and hence is called a \textbf{Lie subgroup} of $\mathbb{G}$.
\end{definition}

\bigskip

\begin{definition}
A \textbf{one parameter subgroup} of a Lie group $\mathbb{G}$ is a smooth
morphism from the additive Lie group of reals $\mathbb{R},+$ to the group
$\mathbb{G}$.
\end{definition}

\bigskip

\subsection{Some examples of Lie groups}

\bigskip

Let $V$ denote an $n$-dimensional vector space over the real numbers
$\mathbb{R}$ with the standard vector inner product which we denote by
$\left\langle \;,\;\right\rangle $.

\begin{itemize}
\item $\mathbb{GL}(n,\mathbb{R})$ The \textbf{real general linear group} of
all automorphisms of the vector space $V$. This can be identified with the
group of all nonsingular $n\times n$ matrices over the reals.

\item $\mathbb{O}(n)$ The \textbf{real orthogonal group} is the group of all
automorphisms which preserve the inner product $\left\langle
\;,\;\right\rangle $. This can be identified with the group of orthogonal
matrices , i.e., matrices $A$ of the form
\[
A^{T}=A^{-1}%
\]
where the superscript $``T"$ denotes the matrix transpose.

\item $\mathbb{SL}(n,\mathbb{R)}$ The \textbf{real special linear group} is
the group of all real $n\times n$ matrices of determinant $1.$
$SL(n,\mathbb{R)}$ is the group of all rigid motions in hyperbolic $n$-space.

\item $\mathbb{SO}(n)=\mathbb{O}(n)\cap\mathbb{SL}(n,\mathbb{R})$ The
\textbf{special orthogonal group} is the group of all orthogonal real $n\times
n$ matrices of determinant $1$.\textbf{\ }This group can be identified with
the group of all rotations in $\mathbb{R}^{n}$ about a fixed point such as the origin.
\end{itemize}

\bigskip\ 

Let $W$ denote an $n$-dimensional vector space over the complex numbers
$\mathbb{C}$ with the standard sesquilinear inner product which we also denote
by $\left\langle \;,\;\right\rangle $.

\begin{itemize}
\item $\mathbb{GL}(n,\mathbb{C})$ The \textbf{complex general linear group} of
all automorphisms of the vector space $W$. This can be identified with the
group of all nonsingular $n\times n$ matrices over the complexes.

\item $\mathbb{SL}(n,\mathbb{C)}$ The \textbf{complex special linear group} is
the group of all complex $n\times n$ matrices of determinant $1.$

\item $\mathbb{U}(n)$ The \textbf{unitary group} is the group of all $n\times
n$ unitary matrices over the complex numbers $\mathbb{C}$, i.e., all $n\times
n$ complex matrices $A$ such that
\[
A^{\dagger}=A^{-1}%
\]
where $A^{\dagger}$ denotes the conjugate transpose.

\item $\mathbb{SU}(n)=\mathbb{U}(n)\cap\mathbb{SL}(n,\mathbb{C)}$ The special
unitary group is the group of all unitary matrices of determinant 1.
\end{itemize}

\bigskip

\subsection{The Lie algebra of a Lie group}

\bigskip

\begin{definition}
Let $\mathbb{G}$ be a Lie group. \ For each element $h\in\mathbb{G}$, we
define the \textbf{right multiplication map}, written $R_{h}$, as
\begin{gather*}
\mathbb{G}\overset{R_{h}}{\longrightarrow}\mathbb{G}\\
g\longmapsto gh
\end{gather*}
The map $R_{h}$ is an autodiffeomorphism of $\mathbb{G}$. \ We let
\[
dR_{h}:T\mathbb{G}\longrightarrow T\mathbb{G}%
\]
denote the corresponding differential of this diffeomorphism.

Finally, a vector field $v\in\mathbf{Vec}(\mathbb{G})$ is said to be
\textbf{right invariant} if
\[
\left(  dR_{h}\right)  \left(  \left.  v\right|  _{g}\right)  =\left.
v\right|  _{gh}%
\]
\end{definition}

\bigskip

\begin{definition}
Let $\mathbf{Vec}_{R}\left(  \mathbb{G}\right)  $ denote the set of right
invariant vector fields on $\mathbb{G}$. \ Then $\mathbf{Vec}_{R}\left(
\mathbb{G}\right)  $ as a subset of the Lie algebra $\mathbf{Vec}\left(
\mathbb{G}\right)  $ inherits the structure of a Lie algebra. \ We call
$\mathbf{Vec}_{R}\left(  \mathbb{G}\right)  $ the \textbf{Lie algebra} of the
Lie group $\mathbb{G}$.
\end{definition}

\bigskip

Let $I$ denote the identity element of the Lie group $\mathbb{G}$. \ Since a
right invariant vector field $v\in\mathbf{Vec}_{R}\left(  \mathbb{G}\right)  $
is completely determined by its restriction to the tangent space
$T_{I}\mathbb{G}$ via
\[
\left.  v\right|  _{g}=\left(  dR_{g}\right)  \left(  \left.  v\right|
_{I}\right)  \text{ ,}%
\]
we can, and do, identify the Lie algebra $\mathbf{Vec}_{R}\left(
\mathbb{G}\right)  $ with the tangent space $T_{I}\mathbb{G}$, i.e.,
\[
\mathbf{Vec}_{R}\left(  \mathbb{G}\right)  =T_{I}\mathbb{G}\text{ .}%
\]

\bigskip

The tangent bundle $T\mathbb{G}$ of a Lie group $\mathbb{G}$ is trivial. \ For
it can be shown that $T\mathbb{G}$ is bundle isomorphic to $\mathbb{G}\times
T_{I}\mathbb{G}$. \ However, there is some additional and useful structure
induced on the Lie algebra $Vec_{R}\left(  \mathbb{G}\right)  =T_{I}%
\mathbb{G}$ by the Lie group structure of $\mathbb{G}$, i.e., the exponential map.

\bigskip

\begin{definition}
We define the \textbf{exponential map} $\exp$ from the Lie algebra
$\mathbf{Vec}_{R}\left(  \mathbb{G}\right)  $ to the Lie group $G$ as
\begin{gather*}
\mathbf{Vec}_{R}\left(  \mathbb{G}\right)  \overset{\exp}{\longrightarrow
}\mathbb{G}\\
\\
v\longmapsto\left.  \left(  e^{vt}\right)  I\right|  _{t=1}%
\end{gather*}
\end{definition}

\bigskip

In other words, we simply follow the flow $\gamma_{v}\left(  t,g\right)
=e^{vt}g$ from the identity $I$ to the point $e^{v}I$ in $\mathbb{G}$.

\bigskip

It can be shown that the exponential map
\[
\exp:\mathbf{Vec}_{R}\left(  \mathbb{G}\right)  \longrightarrow\mathbb{G}%
\]
is a local diffeomorphism. \ It also follows that, for each $v\in
\mathbf{Vec}_{R}\left(  \mathbb{G}\right)  $, $\exp\left(  tv\right)  $ is a
one parameter subgroup of $G$. \ In fact, all one parameter subgroups are of
this form.

\bigskip

\subsection{Some examples of Lie algebras}

\bigskip

\subsubsection{Example: The Lie algebra $\mathbf{u}\left(  N\right)  $ of the
unitary group $\mathbb{U}\left(  N\right)  $.}

\bigskip

In this case, $\mathbf{u}\left(  N\right)  $ is the Lie algebra of all
$N\times N$ skew Hermitian\footnote{A square matrix $M$ is skew Hermitian if
$\overline{M}^{T}=-M$.} matrices of $\mathbb{C}$. \ This can be seen as follows:

\bigskip

The Lie algebra $\mathbf{u}\left(  N\right)  $ is the tangent space
$T^{I}\mathbb{U}\left(  N\right)  $ to $\mathbb{U}\left(  N\right)  $ at the
$N\times N $ identity matrix $I$. \ Hence, $\mathbf{u}\left(  N\right)  $
consists of all tangent vectors $\overset{\bullet}{U}\left(  0\right)
=\left.  \frac{d}{dt}U\left(  t\right)  \right|  _{t=0}$ of all curves
$U\left(  t\right)  $ in $\mathbb{U}\left(  N\right)  $ which pass through $I$
at $t=0$, i.e., which satisfy $U\left(  0\right)  =I$. \ 

\bigskip

Since $U(t)$ is unitary, i.e., since
\[
U\left(  t\right)  \overline{U}\left(  t\right)  ^{T}=I\text{, }%
\]
we find by differentiating the above formula that
\[
\overset{\bullet}{U}\left(  t\right)  \overline{U}\left(  t\right)
^{T}+U\left(  t\right)  \overline{\overset{\bullet}{U}}\left(  t\right)
^{T}=0\text{ .}%
\]
Setting $t=0$, we have
\[
\overline{\overset{\bullet}{U}}\left(  0\right)  ^{T}=-\overset{\bullet}%
{U}\left(  0\right)  \text{ \ .}%
\]
Thus all matrices in $\mathbf{u}\left(  N\right)  $ are skew Hermitian.

\bigskip

Let $M$ be an arbitrary skew $N\times N$ Hermitian matrix. \ Then
\[
U\left(  t\right)  =\exp\left(  tM\right)
\]
is a curve in $\mathbb{U}\left(  N\right)  $ which passes through $I$ at $t=0$
for which $\overset{\bullet}{U}\left(  0\right)  =M$. \ Hence, $\mathbf{u}%
\left(  N\right)  $ is the Lie algebra of all $N\times N$ skew Hermitian
matrices over $\mathbb{C}$. \ 

\bigskip

Let
\[
\sigma_{1}=\left(
\begin{array}
[c]{cc}%
0 & 1\\
1 & 0
\end{array}
\right)  \text{, }\sigma_{2}=\left(
\begin{array}
[c]{cc}%
0 & -i\\
i & 0
\end{array}
\right)  \text{, }\sigma_{3}=\left(
\begin{array}
[c]{cc}%
1 & 0\\
0 & -1
\end{array}
\right)
\]
denote the Pauli spin matrices, and let
\[
\sigma_{0}=\left(
\begin{array}
[c]{cc}%
1 & 0\\
0 & 1
\end{array}
\right)
\]
denote the $2\times2$ identity matrix. \ Then the following is a basis of the
Lie algebra $\mathbf{u}\left(  2^{n}\right)  $
\[
\left\{  \xi_{j_{1}j_{2}\cdots j_{n}}\mid j_{1},j_{2},\ldots,j_{n}%
=0,1,2,3\right\}  \text{ ,}%
\]
where
\[
\xi_{j_{1}j_{2}\cdots j_{n}}=-\frac{i}{2}\sigma_{j_{1}}\otimes\sigma_{j_{2}%
}\otimes\ldots\otimes\sigma_{j_{n}}\text{ \ .}%
\]

\begin{remark}
Please note that, although $\mathbf{u}\left(  N\right)  $ is a Lie algebra of
complex matrices, it is nonetheless a real Lie algebra. \ Thus, the above
basis $\left\{  \xi_{j_{1}j_{2}\cdots j_{n}}\right\}  $ of $\mathbf{u}\left(
2^{n}\right)  $ is a basis of $\mathbf{u}\left(  2^{n}\right)  $ over the
reals $\mathbb{R}$. \ But the matrices in $\mathbf{u}\left(  2^{n}\right)  $
are still matrices of complex numbers!
\end{remark}

\bigskip

\subsubsection{Example: The Lie algebra $\mathbf{su}\left(  N\right)  $ of the
special unitary group $\mathbb{SU}\left(  N\right)  $.}

\bigskip

The Lie algebra $\mathbf{su}\left(  N\right)  $ for the special unitary group
is the same as the Lie algebra of all $N\times N$ traceless skew Hermitian
matrices, i.e., of all $N\times N$ skew Hermitian matrices $M$ such that
$trace\left(  M\right)  =0$. \ A basis of the Lie algebra $\mathbf{su}\left(
2^{n}\right)  $ is
\[
\left\{  \xi_{j_{1}j_{2}\cdots j_{n}}\mid j_{1},j_{2},\ldots,j_{n}%
=0,1,2,3\right\}  -\left\{  \xi_{00\cdots0}\right\}  \text{ \ .}%
\]

\bigskip

\subsubsection{Example: The Lie algebra $\mathbf{so}\left(  3\right)  $ of the
special unitary group $\mathbb{SO}\left(  3\right)  $.}

\bigskip

Finally, we should mention that the Lie algebra $\mathbf{so}\left(  3\right)
$ of the special orthogonal group $\mathbb{SO}\left(  3\right)  $ is the Lie
algebra of all $3\times3$ skew symmetric matrices over the reals $\mathbb{R}$.
\ The following three matrices form a basis for \label{so3 basis} $so\left(
3\right)  $
\[
L_{1}=\left(
\begin{array}
[c]{rrr}%
0 & 0 & 0\\
0 & 0 & -1\\
0 & 1 & 0
\end{array}
\right)  \text{, }L_{2}=\left(
\begin{array}
[c]{rrr}%
0 & 0 & 1\\
0 & 0 & 0\\
-1 & 0 & 0
\end{array}
\right)  \text{, }L_{3}=\left(
\begin{array}
[c]{rrr}%
0 & -1 & 0\\
1 & 0 & 0\\
0 & 0 & 0
\end{array}
\right)  \text{ \ .}%
\]

\bigskip

\subsection{Lie groups as transformation groups on manifolds}

\bigskip

\begin{definition}
Let $M$ be a smooth manifold. \ Then a \textbf{group of transformations acting
on} $M$ is a Lie group $\mathbb{G}$ together with a smooth map
\begin{gather*}
\mathbb{G}\times M\longrightarrow M\\
\left(  g,x\right)  \longmapsto g\cdot x
\end{gather*}
such that

\begin{itemize}
\item[1)] For all $x\in M$, and for all $g_{1},g_{2}\in\mathbb{G}$
\[
g_{1}\cdot(g_{2}\cdot x)=(g_{1}g_{2})\cdot x
\]

\item[2)] For all $\in M$,
\[
e\cdot x=x\text{ ,}%
\]
where $e$ denotes the identity of $\mathbb{G}$.
\end{itemize}

$\mathbb{G}$ is called a transformation group of $M$.
\end{definition}

\bigskip

\begin{definition}
Let $M$ be a smooth manifold, and let $\mathbb{G}$ be a Lie group acting on
$M$. \ Then the action
\[
\mathbb{G}\times M\longrightarrow M
\]
induces an \textbf{infinitesimal action}
\[
\mathbf{Vec}_{R}\left(  \mathbb{G}\right)  \overset{\Psi_{G}}{\longrightarrow
}\mathbf{Vec}(M)\text{ , }%
\]
where $\left.  \Psi_{\mathbb{G}}\left(  v\right)  \right|  _{x}$ is the
tangent vector to the curve
\[
\gamma_{v}\left(  t,x\right)  =e^{tv}x
\]
in $M$ at $x$, i.e.,
\[
\left.  \Psi_{\mathbb{G}}\left(  v\right)  \right|  _{x}=\left.  \frac{d}%
{dt}\left(  e^{tv}x\right)  \right|  _{t=0}\text{ \ .}%
\]
\end{definition}

\bigskip

\subsection{The big and little adjoint representations}

\bigskip

Let $\mathbb{G}$ be a Lie algebra, and let $\mathfrak{g}$ denote the
corresponding Lie algebra.

\bigskip

For each element $h\in\mathbb{G}$, consider the inner automorphism:
\begin{gather*}
\mathbb{G}\overset{\mathcal{I}_{h}}{\longrightarrow}\mathbb{G}\\
g\longmapsto hgh^{-1}%
\end{gather*}
and let
\[
T\mathbb{G}\overset{d\mathcal{I}_{h}}{\longrightarrow}T\mathbb{G}%
\]
denote the corresponding differential. \ We can now define the \textbf{big
adjoint representation }
\[
Ad:\mathbb{G}\longrightarrow Aut(\mathfrak{g})
\]
by
\[
Ad_{h}=\left.  \left(  d\mathcal{I}_{h}\right)  \right|  _{I}%
\]
where $I$ denotes the identity of $G$, and where $Aut(\mathfrak{g})$ denotes
the group of automorphisms of the Lie algebra $\mathfrak{g}$.

\bigskip

We can now in turn define the \textbf{little adjoint representation}
\[
ad:\mathfrak{g}\longrightarrow End\left(  \mathfrak{g}\right)
\]
\ of the Lie algebra $\mathfrak{g}$ by
\[
ad_{v}\left(  u\right)  =\left[  u,v\right]  \text{ ,}%
\]
where $\left[  -,-\right]  $ denotes the Lie bracket, and where $End\left(
\mathfrak{g}\right)  $ denotes the ring of endomorphisms of the Lie algebra
$\mathfrak{g}$.

\bigskip

As the story goes, $End\left(  \mathfrak{g}\right)  $ is actually the Lie
algebra of the Lie group $Aut(\mathfrak{g})$, and we have the following
commutative diagram
\[%
\begin{array}
[c]{ccc}%
\mathfrak{g} & \overset{ad}{\longrightarrow} & End\left(  \mathfrak{g}\right)
\\
\exp\downarrow\qquad &  & \quad\downarrow\exp\\
G & \overset{Ad}{\longrightarrow} & Aut(\mathfrak{g})
\end{array}
\]
which relates the big and little adjoints. \ Little adjoint $ad$ is actually
the differential $d\left(  Ad\right)  $ restricted to the identity $I$ of the
big adjoint $Ad$.

\bigskip

Perhaps the following example would be of help:

\begin{example}
Let $\mathbb{G}$ be the special unitary group $\mathbb{SU}(2)$. \ Let $su(2)$
denote its Lie algebra. \ Then $Aut\left(  \mathbb{G}\right)  $ is the special
orthogonal group $\mathbb{SO}(3)$ and $End\left(  \mathfrak{g}\right)  $ is
the Lie algebra $so(3)$ of $\mathbb{SO}(3)$. \ Thus, we have the familiar
commutative diagram
\[%
\begin{array}
[c]{ccc}%
su(2) & \overset{ad}{\longrightarrow} & so(3)\\
\exp\downarrow\qquad &  & \quad\downarrow\exp\\
\mathbb{SU}(2) & \overset{Ad}{\longrightarrow} & \mathbb{SO}(3)
\end{array}
\]
used in quantum mechanics and in quantum computation.
\end{example}

\bigskip

\begin{remark}
The reader should verify that
\[
ad_{\xi_{j}}=L_{j}%
\]
\end{remark}

\bigskip

\subsection{The orbits of transformation Lie group actions}

\bigskip

Finally, we should remark that the entanglement classes defined previously in
this paper are nothing more than the orbits of a group action. \ For
completeness, we give the definition below:

\bigskip

\begin{definition}
A subset $\mathcal{O}$ of the smooth manifold $M$ is an \textbf{orbit} of the
action of the group $\mathbb{G}$ on $M$ provided

\begin{itemize}
\item[1)] $x\in\mathcal{O}\Longrightarrow g\cdot x\in\mathcal{O}$ for all
$g\in\mathbb{G}$, and

\item[2)] If $S$ is a non-empty subset of $\mathcal{O}$ which satisfies
condition 1) above, then $S=\mathcal{O}$. \ 
\end{itemize}
\end{definition}

\bigskip

In other words, an orbit is a minimal nonempty invariant subset of $M$.


\bigskip

\begin{thebibliography}{9}                                                                                                %

\bibitem {Bell1}Bell, J.S., ``\textbf{Speakable and Unspeakable in Quantum
Mechanics},'' Cambridge University Press (1987).

\bibitem {Bell2}Bell, J.S., Physics, 1, (1964), pp. 3475 - 3467.

\bibitem {Bennett1}Bennett, Charles H., David P. DiVincenzo, Tal Mor, Peter W.
Shor, John A. Smolin, and Barbara M. Terhal, \textbf{Unextendible product
bases and bound entanglement}, quant-ph/9808030.

\bibitem {Bennett2}Bennett, Charles H., David P. DiVincenzo, Christopher A.
Fuchs, Tal Mor, Eric Rains, Peter W. Shor, John A. Smolin, and William K.
Wootters, \textbf{Quantum nonlocality without entanglement, quant-ph/9804053.}

\bibitem {Bennett3}Bennett, Charles H., Herbert J. Bernstein, Sandu Popescu,
and Benjamin Schumacher, \textbf{Concentrating partial entanglement by local
operations}, Phys. Rev. A, Vol. 53, No. 4, April 1996, pp 2046 - 2052.

\bibitem {Brassard1}Brassard, Gilles, Richard Cleve, and Alain Tapp,
\textbf{The cost of exactly simulationg quantum entanglement with classical
communication}, quant-ph/9901035.

\bibitem {Carteret1}Carteret, H.A., and A. Sudbery, \textbf{Local symmetry
properties of 3-qubit states}, quant-ph/0001091.

\bibitem {Carteret2}Carteret, H.A., A. Higuchi, and A. Sudbery,
\textbf{Multipartite generalisation of the Schmidt decomposition}, quant-ph/0006125.

\bibitem {Cerf1}Cerf, Nicholas J. and Chris Adami, ``\textbf{Quantum
information theory of entanglement and measurement},'' in \textbf{Proceedings
of Physics and Computation, PhysComp'96}, edited by J. Leao T. Toffoli, pp 65
- 71. \ See also quant-ph/9605039.

\bibitem {Cox1}Cox, Davis, John Little, and Donal O'Shea, ``\textbf{Ideals,
Varieties, and Algorithms},'' Springer-Verlag (second edition) (1992).

\bibitem {Deutsch2}Deutsch, David, and Patrick Hayden, \textbf{Information
flow in entangled quantum systems}, quant-ph/9906007.

\bibitem {Goodman1}Goodman, Roe, and Nolan R. Wallach,
``\textbf{Representations and Invariants of the Classical Groups},'' Cambridge
University Press (1998).

\bibitem {Einstein1}Einstein, A., B. Podosky, and N. Rosen, \textbf{Can
quantum mechanical description of physical reality be considered complete?},
Phys. Rev. \textbf{47}, 777 (1935); D. Bohm, ``Quantum Theory,''
Prentice-Hall, Englewood Cliffs, NJ (1951).

\bibitem {Eisert1}Eisert, Jens, and Martin Wilkens, \textbf{Catalysis of
entanglement manipulation for mixed states}, quant-ph/9912080.

\bibitem {Englert1}Englert, Berthold-Georg, and Nassaer Metwally,
\textbf{Separability of entangled q-bit pairs}, quant-ph/9912089.

\bibitem {Gruska1}Gruska, Jozef, ``\textbf{Quantum Computing},'' McGraw-Hill, (1999)

\bibitem {Horodecki1}Horodecki,Michal, Pawel Horodecki, Ryszard Horodecki,
\textbf{Limits for entanglement measures}, quant-ph/9908065.

\bibitem {Horodecki2}Horodecki, Pawel, Michal Horodecki, and Ryszard
Horodecki, \textbf{Binding entanglement channels}, quant-ph/9905058.

\bibitem {Horodecki3}Horodecki, Michal, Pawel Horodecki, and Ryszard
Horodecki, \textbf{Separability of n-particle mixed states: necessary and
sufficient conditions in terms of linear maps}, quant-ph/0006071.

\bibitem {Jonathan1}Jonathan, Daniel, and M. Plenio,
\textbf{Entanglement-assisted local manipulation of pure quantum states},
Phys. Rev. Lett. 83, 3566 (1999). (quant-ph/9905071)

\bibitem {Jonathan2}Jonathan, Daniel, and Martin B. Plenio,
\textbf{Entanglement-assisted local manipulation of pure quantum states,} quant-ph/9905071.

\bibitem {Kus1}Kus, Marek, and Karol Zyczkowski, \textbf{Geometry of entangled
states}, quant-ph/0006068.

\bibitem {Lewenstein1}Lewenstein, M., D. Bruss, J.I. Cirac, B. Kraus, M. Kus,
J. Samsonowicz, A. Sanpera, and R. Tarrach, \textbf{Separability and
distillability in composite quantum systems - a primer -,} quant-ph/0006064.

\bibitem {Linden1}Linden, N., and S. Popescu, \textbf{On multi-particle
entanglement}, quant-ph/9711016.

\bibitem {Linden2}Linden, N., S. Popescu, and A. Sudbery, \textbf{Non-local
properties of multi-particle density matrices}, quant-ph/9801076.

\bibitem {Linden3}Linden, N. and Sandu Popescu, \textbf{Good dynamics versus
bad kinematics. \ Is entanglement needed for quantum computation?}, quant-ph/9906008.

\bibitem {Lo1}Lo, Hoi-Kwong, Sandu Popescu, and Tim Spiller, ``Introduction to
Quantum Computation and Information,'' World Scientific (1998).

\bibitem {Lomonaco1}Lomonaco, Samuel J., Jr., \textbf{A Rosetta stone for
quantum mechanics with an Introduction to Quantum Computation: Lecture Notes
for the AMS Short Course on Quantum Computation, Washington, DC, January
2000},'' to appear in the AMS\ PSAMP Series. (Quant-Ph/0007045)

\bibitem {Lomonaco2}Lomonaco, Samuel J., Jr., \textbf{The Shor/Simon algorithm
from the perspective of group representation theory}, to appear in
``\textbf{Quantum Computation and Information},'' AMS\ Contemporary
Mathematics Series (2001).

\bibitem {Makhlin1}Makhlin, Yuriy, \textbf{Nonlocal properties of two-qubit
gates and mixed states and optimization of quantum computations}, quant-ph/0002045.

\bibitem {Nielsen1}Nielsen, Michael A., \textbf{Conditions for a class of
entanglement transformations}, Physical Review Letters, Vol 83 (2), pp
436--439 (1999). (quant-ph/981105)

\bibitem {Nielsen2}Nielsen, Michael A., \textbf{Majorization and its
applications to quantum information theory}, preprint.

\bibitem {Nielsen3}Nielsen, M.A., \textbf{Continuity bounds for entanglement},
Phys. Rev. A, Vol. 61, (2000)

\bibitem {Nielsen4}Nielsen, M.A., \textbf{Characterizing mixing and
measurement in quantum mechanics}, quant-ph/0008073.

\bibitem {O'Connor1}O'Connor, Kevin M., anf William K. Wootters,
\textbf{Entangled rings}, quant-ph/0009041.

\bibitem {Nielsen5}Nielsen, Michael A., and Isaac L. Chuang, \textbf{``Quantum
Computation and Quantum Information,'' Cambridge University Press (2000).}

\bibitem {Olver1}Olver, Peter J., \textbf{``Applications of Lie Groups to
Differential Equantions,''} Springer-Verlag, (1993).

\bibitem {Pontrjagin1}Pontrjagin, Leon, \textbf{``Topological Groups,''}
Princeton University Press, (1958).

\bibitem {Rains1}Rains, Eric M., \textbf{Polynomial invariants of quantum
codes}, quant-ph/9704042.

\bibitem {Sattinger1}Sattinger, D.H., and O.L. Weaver, ``\textbf{Lie Groups
and Algebras with Applications to Physics, Geometry, and Mechanics},''
Springer-Verlag, (1993).

\bibitem {Schlienz1}Schlienz, J., and G. Mahler, Physics Letters A 39 (1996).

\bibitem {Shor1}Shor, Peter W., \textbf{Polynomial time algorithms for prime
factorization and discrete logarithms on a quantum computer}, SIAM\ J.
Computing, 26(5) (1997). pp 1484 - 1509.

\bibitem {Shor2}Shor, Peter W., John A. Smolin, and Ashish V. Thapliyal,
\textbf{Superactivation of bound entanglement}, quant-ph/0005117.

\bibitem {Spivak1}Spivak, Michael, ``\textbf{A Comprehensive Introduction to
Differential Geometry},'' Volumes 1-5, Publish or Perish, Inc. (1979).

\bibitem {Sudbery1}Sudbery, Anthony, \textbf{On local invariants of
three-qubit states}, quant-ph/0001116.

\bibitem {Sudbery2}Sudbery, Anthony, \textbf{The space of local equivalence
classes of mixed two-qubit states}, quant-ph/0001115.

\bibitem {Terhal1}Terhal, Barbara M., and Pawel Horodecki, \textbf{A Schmidt
number for density matrices}, quant-ph/9911117.

\bibitem {Virmani1}Virmani, S., and M.B. Plenio, \textbf{Ordering states with
entanglement measures}, quant-ph/9911119.

\bibitem {Wallach1}Wallach, N.R., and J. Willenbring, \textbf{On some }%
$q$-\textbf{analogs of a theorem of Kostant-Rallis}, Canad. J. Math., Vol.
\textbf{52} (2), 2000, pp. 438-448.

\bibitem {Warner1}Warner, Frank W., ``\textbf{Foundations of Differential
Manifolds and Lie Groups},'' Scott, Foresman and Company, Glenview, Illinois, (1971).
\end{thebibliography}
\end{document}